\documentclass[pre,10pt,twocolumn,superscriptaddress]{revtex4-1}

\usepackage{mathtools,amssymb,bm}

\usepackage{soul}

\usepackage{mathtools}
\newcommand{\myrightleftarrows}[1]{\mathrel{\substack{\xrightarrow{#1} \\[-.9ex] \xleftarrow{#1}}}}

\newcommand{\eref}[1]{Eq.\,(\ref{#1})}
\newcommand{\fref}[1]{Fig.\,\ref{#1}}
\newcommand{\tref}[1]{Table\,\ref{#1}}
\newcommand{\sref}[1]{Section\,\ref{#1}}
\newcommand{\Aref}[1]{Appendix\,\ref{#1}}

\usepackage{calc}

\newcommand*\settablecounter[2]{%
        \setcounter{table}{#2-1}%
}

\newcommand{\ignore}[1]{} 

\begin{document}

\title{ Demographic noise can reverse the direction of deterministic selection }

\author{George W. A. Constable}
\affiliation{Department of Ecology and Evolutionary Biology, Princeton University, NJ 08544, U.S.A.}
\author{Tim Rogers}
\affiliation{Department of Mathematical Sciences, University of Bath, BA2 7AY, U.K.},
\author{Alan J. McKane}
\affiliation{School of Physics and Astronomy, The University of Manchester, M13 9PL, U.K.}
\author{Corina E. Tarnita}
\affiliation{Department of Ecology and Evolutionary Biology, Princeton University, NJ 08544, U.S.A.}




\begin{abstract}
{Deterministic evolutionary theory robustly predicts that populations displaying altruistic behaviors will be driven to extinction by mutant “cheats” that absorb common benefits but do not themselves contribute. Here we show that when demographic stochasticity is accounted for, selection can in fact act in the reverse direction to that predicted deterministically, instead favoring cooperative behaviors that appreciably increase the carrying capacity of the population. Populations that exist in larger numbers experience a selective advantage by being more stochastically robust to invasions than smaller populations, and this advantage can persist even in the presence of reproductive costs. We investigate this general effect in the specific context of public goods production and find conditions for stochastic selection reversal leading to the success of public good producers. This insight, developed here analytically, is missed by both the deterministic analysis as well as standard game theoretic models that enforce a fixed population size. The effect is found to be amplified by space; in this scenario we find that selection reversal occurs within biologically reasonable parameter regimes for microbial populations. Beyond the public good problem, we formulate a general mathematical framework for models that may exhibit stochastic selection reversal. In this context, we describe a stochastic analogue to $r-K$ theory, by which small populations can evolve to higher densities in the absence of disturbance.}
\end{abstract}


\maketitle

Over the past century, mathematical biology has provided a framework with which to begin to understand the complexities of evolution. Historically, development has focused on deterministic models~\cite{hofbauer_1998}. However, when it comes to questions of invasion and migration in ecological systems, it is widely acknowledged that stochastic effects may be paramount, since the incoming number of individuals is typically small. The importance of demographic (intrinsic) noise has long been argued in population genetics; it is the driver of genetic drift and can undermine the effect of selection in small populations~\cite{fisher_1930,wright_1931}. This concept has also found favor in game theoretic models of evolution which seek to understand how apparently altruistic traits can invade and establish in populations~\cite{nowak_2006}. However, the last decade has seen an increase in the awareness of some of the more exotic and counter-intuitive aspects of demographic noise: it has the capacity to induce cycling of species~\cite{mckane_2005}, pattern formation~\cite{butler_2009,hallatschek_2007}, speciation~\cite{rossberg_2013} and spontaneous organization in systems that do not display such behavior deterministically.

Here we explore the impact of demographic noise on the direction of selection in interactions between multiple phenotypes or species. Historically, a key obstacle to progress in this area has been the analytical intractability of multidimensional stochastic models. This is particularly apparent when trying to investigate problems related to invasion, where systems are typically far from equilibrium. A promising avenue of analysis has recently become apparent however through stochastic fast-variable elimination~\cite{parsons_quince_2007_1,doering_2012}. If a system consists of processes that act over very different timescales, it is often possible to eliminate fast-modes, assumed to equilibrate quickly in the multidimensional model, and obtain a reduced dimensional description that is amenable to analysis~\cite{gunawardena_2014}. This approach has been employed multiple times over the last decade to study a stochastic formulation of the classical Lotka-Volterra competition model for two competing phenotypes/species. In \cite{parsons_quince_2007_1,parsons_quince_2007_2,doering_2012,nelson_2015,constable_2015}, such models were analyzed under the assumption that the dynamics regulating the total population size (birth, death and competition) occurred on a much faster timescale than the change in population composition. In particular \cite{parsons_quince_2007_1,parsons_quince_2007_2,doering_2012,nelson_2015} have shown that it is possible for systems that appear neutral in a deterministic setting to become non-neutral once stochasticity is included. If the two phenotypes have equal deterministic fitness, but one is subject to a larger amount of demographic noise than the other, then the effect of this noise alone can induce a selective drift in favor of the phenotype experiencing less noise. This stems from the fact that it is easier to invade a noisy population than a stable one; furthermore, the direction of this induced selection can vary with the system's state~\cite{parsons_quince_2010}. The idea has been further generalized mathematically in \cite{kogan_2014}. 

Here we will show more generally that not only can stochasticity break deterministic neutrality, but that it has the capacity to reverse the direction of selection predicted deterministically. Thus while in a deterministic setting a certain phenotype will always reach fixation (and is resistant to invasions), in a stochastic setting its counterpart can in fact be more likely to invade and fixate (and less susceptible to invasions). These results generalize recent work on modified Moran and Wright-Fisher type models \cite{houch_2012,houch_2014} to a large class of models consisting of two phenotypes interacting with their environment. We begin with the analysis of a prototypical public good model, which is used to illustrate our analysis. We find that stochastic selection reversal can alleviate the public good production dilemma. We further show how space can amplify this phenomenon, allowing the reversal of selection to emerge over a greater parameter range. Finally, we extend the ideas to a more general model framework, and explore the types of system in which we expect this behavior to be relevant. In particular we discuss the similarities with $r-K$ selection theory~\cite{reznick_2002}.

\section{Public Good model}

It is generally accepted that random events play a strong role in the evolution of cooperative behavior, which is deterministically selected against~\cite{nowak_2006}. The standard formulation of evolutionary game theory involves setting the problem in terms of a modified Moran model~\cite{nowak_2004,rice_evo_theory}. The Moran model is a population genetic model first developed as an abstract illustration of the effect of genetic drift in a haploid population of two phenotypes; an individual is picked to reproduce with a probability proportional to their fitness, whilst simultaneously a second individual is chosen randomly to die~\cite{crow_kimura_into}. Coupling birth and death events keeps the population size fixed, which increases the tractability of the system.

The specification of fixed population size is however restrictive and can be problematic. Most prominently, a phenotype with increased fitness can be no more abundant in isolation than its ailing counterpart. Additional difficulties are encountered if one attempts to use simple game-theoretic models to quantitatively understand more complex experimental data. While, for example, assuming some arbitrary non-linearity in the model's game payoff matrix may enable experimental findings to be elegantly recapitulated, it is more difficult to justify the origin of these assumptions on a mechanistic level~\cite{gore_2009}. In light of such issues, it has been suggested that a more ecologically grounded take on the dynamics of cooperation might be preferable~\cite{hauert_2006,huang_2015}, one in which the population size is not fixed and that is sufficiently detailed that mechanistic (rather than phenomenological) parameters can be inferred experimentally. In the following, we take such an approach. We begin by considering a prototypical model of public good production and consumption.

In our model, we consider a phenotype $X$ having the ability to produce a public good $Q$ that catalyzes its growth. We wish to capture the stochastic dynamics of the system. To this end we assume that the system is described by a set of probability transition rates, which describe the probability per unit time of each reaction occurring:
\begin{eqnarray}
&  X \overset{b_{x}}{\underset{\kappa/R^{2}}{ \myrightleftarrows{\rule{0.4cm}{0cm}} } } X + X \,,  \quad &  X + Q \xrightarrow{ r / R^{2} } X + X + Q  \,,   \nonumber \\
& X\xrightarrow{ p_{x} } X + Q \,,\quad & Q\xrightarrow{\delta}\varnothing\,. \label{eq_reactions_1}
\end{eqnarray}
In the absence of the public good, the producer phenotype $X$ reproduces at a baseline birthrate $b_{x}$. The phenotypes encounter each other and the public good at a rate $R^{-2}$; the quantity $R^{2}$ can be interpreted as a measure of the area (or volume) to which the system is confined. Death of the phenotype occurs solely due to crowding effects at rate $\kappa$, multiplied by the encounter rate. Phenotypes encounter and utilize the public good at a rate $r/R^{2}$. We study the case where this reaction is catalytic (i.e. the public good is conserved) and leads to a phenotype reproduction. Examples of catalytic (reusable) public goods are the enzyme invertase produced by the yeast \textit{Saccharomyces cerevisiae}~\cite{koschwanez_2011} or the siderophore pyoverdine produced by the bacterium \textit{Pseudomonas aeruginosa}~\cite{kummerli_2010}. The total rate at which the phenotype reproduces is thus increased in the presence of the public good. The public good itself is produced by the producer phenotype at a rate $p_{x}$ and decays at a rate $\delta$.  Note that as well as controlling the spatial scale of the well-mixed system, the magnitude of $R$ will also control the typical number of individuals in the system, since larger $R$ (more space) allows the population to grow to greater numbers. We next introduce a mutant phenotype $Y$ that does not produce the public good; (i.e. $p_{y}=0$) consequently, it has a different baseline birth rate $b_{y}$ which we expect to be at least as high as that of the producer, due to the non-producers' reduced metabolic expenditure. Its interactions with the public good are otherwise similar to those of $X$ (see \eref{eq_reactions_1}).

The state of the system is specified by the discrete variables $n_{x}$, $n_{y}$ and $n_{q}$, the number of each phenotype and public good respectively. For the system described, we wish to know the probability of being in any given state at any given time. To answer this, we set up an infinite set of partial difference equations (one for each unique state $(n_x,n_y,n_q)$) that measures the flow of probability between neighboring states (controlled by the transitions \eref{eq_reactions_1}). These equations govern the time-evolution of a probability density function $P(n_{x},n_{y},n_{q},t)$ (see \eref{eq:supmat_meqn_general}). Such a model is sometimes termed a microscopic description~\cite{gardiner_2009}, since it takes account of the dynamics of discrete interactions between the system variables. 

Although the probabilistic model is straightforward to formalize, it is difficult to solve in its entirety. We apply an approximation that makes the model more tractable, while maintaining the system's probabilistic nature. Such approximations, which assume that the system under consideration has a large but finite number of individuals, are well practiced and understood~\cite{gardiner_2009} and are analogous to the diffusion approximation~\cite{crow_kimura_into} of population genetics. Assuming that $R$ is large, but finite, (which implies a large number of individuals in the system), we transform the system into the approximately continuous variables $(x,y,q)=(n_{x},n_{y},n_{q})/R^{2}$ and expand the partial difference equations in $1/R^{2}$. This allows us to to express the infinite set of partial difference equations as a single partial differential equation in four continuous variables, $(x,y,q,t)$. However, since the PDE results from a Taylor expansion, it has infinite order. Truncating the expression after the first term (at order $R^{-2}$), one obtains a deterministic approximation of the dynamics (valid for $R \rightarrow \infty$, or equivalently for infinite population sizes). Since we aim to make the system tractable but still retain some stochastic element to the dynamics, we truncate the expansion after the second term (at order $R^{-4}$, see \eref{eq:supmat_eq_FPE_general}). The resulting model can be conveniently expressed as a set of It\={o} stochastic differential equations (SDEs):
\begin{eqnarray}
\dot x &=& x \left[ b_{x} + r q -\kappa (x+y) \right] + R^{-1} \eta_{x}(t)\,, \nonumber \\
\dot y &=& y \left[ b_{y} + r q -\kappa (x+y) \right] + R^{-1}  \eta_{y}(t)\,, \label{eq_PG_SDE} \\
\dot q &=& p_{x} x - \delta q +  R^{-1}  \eta_{q}(t) \,. \nonumber
\end{eqnarray}
The $\eta_{i}(t)$ represent Gaussian white noise terms whose correlations depend on the state of the system (the noise is multiplicative). Importantly, because \eref{eq_PG_SDE} has been developed as a rigorous approximation of the underlying stochastic model, \eref{eq_reactions_1}, the precise functional form of the noise can be determined explicitly, rather than posited on an ad-hoc basis (see \Aref{sec:supmat_sec_obtain_SDE_system}). Setting $ R \rightarrow \infty $, the population size increases with the interaction scale and one recovers the deterministic limit. Since \eref{eq_PG_SDE} is a course-grained approximation of the underlying microscopic model but retains an inherent stochasticity, it is often referred to as the mesoscopic limit~\cite{mckane_TREE}.

First we analyze the dynamics of \eref{eq_PG_SDE} in the deterministic, $ R \rightarrow \infty $ limit. There exist three fixed points, or equilibria. The first, at the origin, is always unstable. The remaining fixed points occur when the system only contains a single phenotype: the producer fixed point, $(x,y,q)=( K_{x} , 0 , p_{x} K_{x}/\delta )$ and the non-producer fixed point, $(x,y,q)=( 0 , K_{y} , 0 )$. Thus $K_{x}$ and $K_{y}$ are measures of the phenotypes' frequency (carrying capacity) in isolation, with precise forms
\begin{equation}
K_{x}= \frac{ b_{x} \delta }{ \kappa \delta - p_{x} r }\,, \qquad K_{y}= \frac{ b_{y}  }{ \kappa  }\,.  \label{eq_carrying_capacity}
\end{equation}
If $b_y>b_x$ then the non-producer fixed point is always stable while the producer fixed point is always unstable. However, the non-producer fixed point is only globally attracting if $ \kappa \delta > r p_{x} $. If this condition is not met then there exist initial conditions for which the producers produce and process the public good faster than they die and faster than the public good degrades, resulting in unbounded exponential growth of the system. This biologically unrealistic behavior comes from the fact that we have assumed for simplicity that the public good uptake does not saturate. Since this behavior is unrealistic, we will work in the regime $ \kappa \delta > r p_{x} $ for the remainder of the paper. Finally, we are interested in systems where the size of the producer population in isolation is larger than that of the non-producer, $K_x>K_y$; this is true if the condition $b_{x}>b_{y}(1-r p_x/\delta\kappa)$ holds. Thus, deterministically, a non-producing mutant will always take over a producer population and, due to the absence of the public good, it will yield a smaller population at equilibrium. 

This deterministic analysis predicts, unsurprisingly, that a population composed entirely of non-producers is the only stable state. We next explore the behavior of the system in \eref{eq_reactions_1} when demographic stochasticity is considered.

\subsection{Mesoscopic selection reversal}

Due to noise, a stochastic system will not be positioned precisely on deterministic fixed points, but rather it will fluctuate around them. In the above system, these fluctuations will occur along the $y$-axis for the non-producer fixed point while in the absence of non-producers they will occur in the $(x,q)$ plane for the producer fixed point. We can define $N_{x} = R^{2} K_{x}$ and $N_{y} = R^{2} K_{y}$ to be the mean number of the phenotypes $X$ and $Y$ in isolation in the respective stationary states. We assume that the non-producing phenotype has a greater per-capita birth rate than the producer phenotype, i.e. $b_{y} > b_{x}$, and we introduce a single non-producing mutant into a producer population. While the deterministic theory predicts that the non-producer should sweep through the population until it reaches fixation, in the stochastic setting fixation of the non-producer is by no means guaranteed: there is a high probability that the single mutant might be lost due to demographic noise. However, since the non-producer is deterministically selected for, we might expect the probability of a non-producer mutant invading and fixating in a resident producer population to be greater than the probability of a producer mutant invading and fixating in a resident non-producer population. We will explore this question below. 

In order to make analytic predictions about the stochastic model, we need to reduce the complexity of the system. This can be done if we employ methods based on the elimination of fast variables~\cite{parsons_rogers_2015} to obtain an effective one-dimensional description of the system dynamics. To this end, we begin by assuming that the public good production and decay, $p_{x}$ and $\delta$, and the phenotypes' reproduction and death, $b_{x}$, $b_{y}$, and $\kappa$, occur on a much faster timescale than the rate of change of population composition, which is governed by the difference in birth rates, $b_x - b_y$. Essentially this assumes that the cost of public good production is marginal. In the case of \textit{S. cerevisiae}, this assumption is supported by empirical work (see \tref{tab:supmat_fig_parameter_list_1}). In order to mathematically investigate this timescale-separation we define
\begin{equation}
 b_{x} = b(1-\varepsilon)\,,\quad b_{y} = b \,, \label{eq_def_b}
\end{equation}
where the parameter $\varepsilon$ represents the metabolic cost that $X$ pays for producing the public good. The parameter $\varepsilon$ now controls the rate of change of population composition, and if $1>>\varepsilon$, we have our desired timescale separation in the deterministic system. Because the parameters $K_{x}$, $K_{y}$, $N_{x}$ and $N_{y}$ depend on $\varepsilon$, we will find it convenient to define their values when $\varepsilon=0$ as $K_{x}^{(0)}$, $K_{y}^{(0)}$, $N_{x}^{(0)}$ and $N_{y}^{(0)}$ respectively. In order to maintain our assumption that the composition of the phenotype population changes slowly in the stochastic system, we additionally require that the noise is small. However this assumption has already been implicitly made in the derivation of \eref{eq_PG_SDE},  where it was assumed that $R$ is large, and thus $R^{-1}$, the prefactor for the noise terms, is small. In order to formalize this, we will find it convenient to assume $R^{-2}\approx \mathcal{O}\left(  \varepsilon \right)$.

\begin{figure}[t]
\setlength{\abovecaptionskip}{-2pt plus 3pt minus 2pt}
\begin{center}
\includegraphics[width=0.5\textwidth]{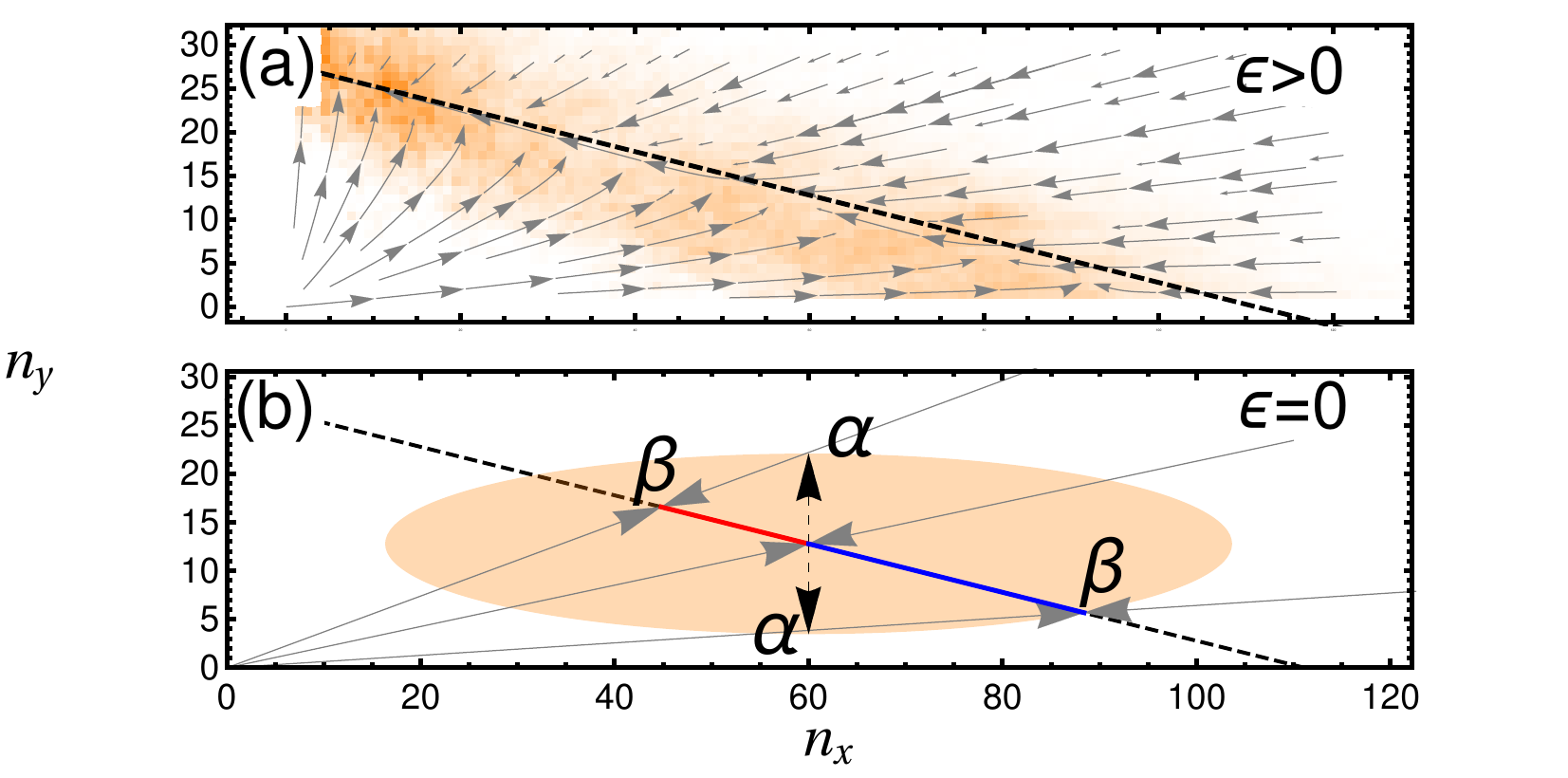}
\end{center}
\caption{System dynamics in the phenotype plane. Deterministic trajectories shown as gray arrows. Panel (a): Trajectories rapidly collapse to a SM (black dashed line), before slowly moving to the non-producing $Y$ fixed-point. Stochastic trajectories (histogram overlaid in orange) remain in the region of the SM but may fluctuate away from it. Panel (b): Illustration of the origin of noise-induced selection. The orange ellipse depicts the standard deviation of Gaussian fluctuations originating at its center. Fluctuations (black dashed arrows) to points $\alpha$ are equally likely, however when projected back to the CM (black dashed line) to points $\beta$, a bias for producing $X$ phenotype is observed. Parameters used are $p_{x}=9.5\times10^{-4}$, $\varepsilon=0.08$ in panel (a), $\varepsilon=0$ in panel (b) and the remaining parameters are given in \tref{tab:supmat_fig_parameter_list_1}. }\label{fig_streamplots}
\end{figure}

Under the above assumptions, the system features a separation of timescales. Next, we take advantage of this to reduce the complexity of the system. Deterministically, the existence of a set of fast timescales suggests the existence of a lower-dimensional subspace, the slow manifold (SM), shown in \fref{fig_streamplots}(a), to which the system quickly relaxes, and along which it slowly moves, until it reaches the system's stable fixed point. This behavior can be exploited if we assume that the system reaches the SM instantaneously. We can then describe the dynamics of the entire system in this lower dimensional space, and thus reduce the number of variables in our description of the deterministic system. However, we are interested in the stochastic dynamics.

The stochastic trajectories initially collapse to the region around the SM, about which they are confined, but along which they can move freely until one of the phenotypes fixates (see \fref{fig_streamplots}(a)). Fluctuations that take the system off the SM are quickly quashed back to another point on the SM; however the average position on the SM to which a fluctuation returns is not necessarily the same as that from which the fluctuation originated. A crucial element of the dynamics in this stochastic setting is that the form of the noise, combined with that of the trajectories back to the SM, can induce a bias in the dynamics along the SM (see \fref{fig_streamplots}(b)). This is the origin of the stochastic selection reversal that we will explore. In order to capture this behavior while simultaneously removing the fast timescales in the stochastic system, we map all fluctuations off the SM along deterministic trajectories back to the SM~\cite{parsons_rogers_2015}. This essentially assumes that any noisy event that takes the system off the SM is instantaneously projected back to another point on the SM.

For clarity, we briefly describe the dynamics when $\varepsilon=0$. In this case the birth rates of phenotypes $X$ and $Y$ are identical. Instead of the two non-zero fixed points, $K_{x}$ and $K_{y}$, found above, the deterministic system now has a line of fixed points, referred to as a center manifold (CM)~\cite{arnold_2003}. The CM is identical to the SM in the limit $\varepsilon\rightarrow0$. It is given by 
\begin{equation}
 y = \frac{ K_{y}^{(0)} }{ K_{x}^{(0)} } \left( K_{x}^{(0)} - x \right) \,, \quad q = \frac{ p_{x} }{ \delta } x \,, \label{eq_CM}
\end{equation}
and shown graphically in \fref{fig_streamplots}(b). The separation of timescales in the system is now at its most pronounced, since there are strictly no deterministic dynamics along the CM following the fast transient to the CM. However the stochastic system still features dynamics along the CM. Applying the procedure outlined in \cite{parsons_rogers_2015} we arrive at a description of the stochastic dynamics in a single variable, the frequency of producers along the CM;
\begin{equation}
 \dot{x} = \frac{ b }{ R^{2} } x \left( 1 - \frac{ x }{ K_{x}^{(0)} } \right) \,\mathcal{F}(x)   + \frac{1}{R} \zeta(t) \,, \label{eq_reduced_PG_SDE_neutral} 
\end{equation}
where
\begin{equation}
 \mathcal{F}(x) = 2 \left( \frac{ K_{x}^{(0)} - K_{y}^{(0)} }{ (K_{x}^{(0)}K_{y}^{(0)})^{2} } \right) \left[ K_{x}^{(0)} K_{y}^{(0)} + \left( K_{x}^{(0)} - K_{y}^{(0)} \right) x \right] \,. \nonumber
\end{equation}
Here $\zeta(t)$ is a Gaussian white noise term with a correlation structure given in \eref{eq:supmat_eq_PG_clean_Bbar}. Together with \eref{eq_CM}, \eref{eq_reduced_PG_SDE_neutral} approximates the dynamics of the entire system. Note that while \eref{eq_reduced_PG_SDE_neutral} predicts a noise-induced directional drift along the CM (controlled by $\mathcal{F}(x)$), a deterministic analysis predicts no dynamics, since the CM is by definition a line of fixed points. This directional drift along the CM results from the projection bias illustrated in \fref{fig_streamplots}(b). If $p_{x}>0$, then $K_{x}^{(0)}>K_{y}^{(0)}$, and so $\mathcal{F}(x) > 0 $; thus the public good production by phenotype $X$ induces a selective pressure that selects for $X$ along the center manifold.

The origin of the term $\mathcal{F}(x)$ in \eref{eq_reduced_PG_SDE_neutral} can be understood more fully by exploring its implications for the invasion probabilities of $X$ and $Y$, denoted $\phi_{x}$ and $\phi_{y}$. These can be straightforwardly calculated since the system is one-dimensional (see \Aref{sec:supmat_sec_fixprob}). We find 
\begin{equation}
 \phi_{x} = \frac{1}{N_{y}} \,, \qquad \textrm{and} \qquad \phi_{y} = \frac{1}{N_{x}} \,, \label{eq_phi_quasineutral}
\end{equation}
where $\phi_{x}>\phi_{y}$ so long as $p_{x} > 0$ (see \eref{eq_carrying_capacity}). The term $\mathcal{F}(x)$ can thus be interpreted as resulting from the stochastic advantage the producers have at the population level from reaching higher carrying capacities in isolation, which makes them more stochastically robust to invasion attempts. This result is independent of the spatial scale $R$ (and therefore population size) so long as $R$ is finite. 

If $\varepsilon \neq 0 $, the system does not collapse to the CM, but rather to the SM. At leading order in $\varepsilon$, the equation for the SM is given by \eref{eq_CM}. Upon removing the fast dynamics, the effective dynamics of $x$ can now be shown to take the form (see \eref{eq:supmat_eq_PG_maintext_effective})
\begin{equation}
 \dot{x} =  b x \left( 1 - \frac{x}{K_{x}^{(0)}} \right) \left( \frac{ 1 }{ R^{2} } \,\mathcal{F}(x) - \varepsilon   \right)  + \frac{1}{R} \zeta(t) \,, \label{eq_reduced_PG_SDE}
\end{equation}
where $\zeta(t)$ and $\mathcal{F}(x)$ are the same as in \eref{eq_reduced_PG_SDE_neutral}. The SDE now consists of two components. The deterministic contribution, governed by $\varepsilon$, exerts a selective pressure against phenotype $X$, due to its reduced birth rate. The stochastic term, $\mathcal{F}(x)$ exerts a pressure in favor of phenotype $X$, resulting, as in the case $\varepsilon = 0$ discussed above, from the producers’ stochastic robustness to invasions.

\begin{figure}[t]
\setlength{\abovecaptionskip}{-2pt plus 3pt minus 2pt}
\begin{center}
\includegraphics[width=0.5\textwidth]{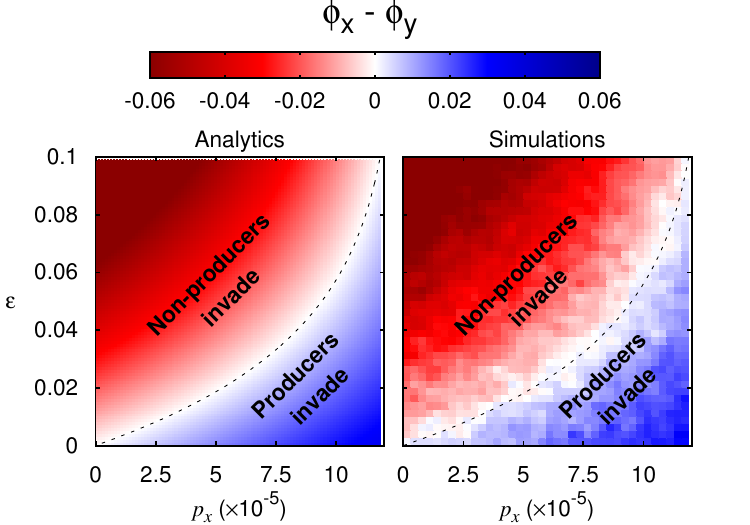}
\end{center}
\caption{Stochasticity can render non-producers more susceptible to invasion by producers than vice versa. Plots of the difference in invasion probabilities between producers $X$ and non-producers $Y$ as a function of the cost to birth for production, $\varepsilon$, and good production rate $p_{x}$. The remaining parameters are taken from  \tref{tab:supmat_fig_parameter_list_1}. Left: analytic results for a single small patch (see \eref{eq_eps_condition}). The critical cost $\varepsilon$ for selection reversal, \eref{eq_eps_condition}, is shown here as the black dashed line. Right; results from Gillespie simulations~\cite{gillespie_1976} of the stochastic process \eref{eq_reactions_1}, averaged over $2000$ runs.}\label{fig_trade_off}
\end{figure}

Thus, when $\varepsilon > 0$, a trade-off emerges in the stochastic system between the stochastic advantage to public good production (due to increased population sizes) and the deterministic cost producers pay (in terms of birth rates). If the birth costs are not too high, producers will be selected for, which constitutes a reversal in the direction of selection from the deterministic prediction. Specifically, we can calculate the condition on the metabolic cost that ensures that the producers are fitter than the non-producers (i.e. $\phi_{x}>\phi_{y}$):
\begin{equation}
 \varepsilon < \frac{\kappa}{ b R^{2} } \log  \left[ \frac{\delta  \kappa }{ \delta \kappa - p_{x} r } \right]  \,.   \label{eq_eps_condition}
\end{equation}
Whereas for no metabolic cost producers consistently have a stochastic advantage regardless of typical population size (see \eref{eq_phi_quasineutral}), for non-zero production costs, the population must be sufficiently small that stochastic effects, governed by $R^{-2}$, are dominant. \fref{fig_trade_off} shows that the theory predicts well the trade-off in the underlying stochastic model (\ref{eq_reactions_1}).

We have shown that stochastic selection reversal is more prevalent when $R$ is not large. Meanwhile our analytic results results have been obtained under the assumption that $R$ is large, which allowed us to utilize the diffusion approximation leading to \eref{eq_PG_SDE} and aided the timescale elimination procedure that yielded \eref{eq_reduced_PG_SDE}. We therefore expect that although stochastic selection reversal will become more prominent as $R$ is reduced, the quality of our analytic predictions may suffer. Despite this caveat, it is the small $R$ regime that is interesting to us. Small values of $R$ are associated with small population sizes. While it is conceivable that populations of macro-organisms may consist of a small number of individuals, this limit is not so pertinent to the study of micro-organisms. In the next section however, we will show that by incorporating space, the constraint of small population size can be relaxed.

\section{Spatial amplification}

In this section we consider a metapopulation on a grid: each subpopulation (patch) has a small size so that demographic noise continues to be relevant locally, but the number of subpopulations is large so that the overall population in the system is large. This method of incorporating demographic stochasticity into spatial systems has proved to be successful in the modeling of microbial populations~\cite{hallatschek_2007}. We consider a grid of $C$ patches. The dynamics within each patch are given by the transitions in \eref{eq_reactions_1}, and coupled to the surrounding patches by the movement of the phenotypes and public good. A patch will produce migrants at a rate proportional to its density. Producers $X$ and non-producers $Y$ disperse with a probability rate $m$ to a surrounding region, while the public good diffuses into neighboring regions at a rate $D$. Once again the diffusion approximation can be applied to obtain a set of SDEs approximating the system dynamics;  
\begin{eqnarray}
\frac{\mathrm{d}x_{ij}}{\mathrm{d}\tau} &=& x_{ij} \left( b_{x} + r q_{ij}  - \kappa ( x_{ij} + y_{ij} ) \right) + m \left(  L \bm{x} \right)_{ij} + \frac{ \eta_{xij}(t) }{ R } \, , \nonumber \\
\frac{\mathrm{d}y_{ij}}{\mathrm{d}\tau} &=& y_{ij} \left( b_{y} + r q_{ij}  - \kappa ( x_{ij} + y_{ij} ) \right) + m \left(  L \bm{y} \right)_{ij} + \frac{ \eta_{yij}(t) }{ R }\, ,  \nonumber \\
\frac{\mathrm{d}q_{ij}}{\mathrm{d}\tau} &=& p_{x} x_{ij} - \delta q + D \left( L \bm{q} \right)_{ij} + \frac{ \eta_{qij}(t) }{ R } \,,
\label{eq_spatial_ODEs} 
\end{eqnarray} 
where $ij$ is the patch on row $i$ and column $j$. The operator $L$ is the discrete Laplacian operator $(L\bm{x})_{ij} = - 4 x_{ij} + x_{(i-1)j} + x_{(i+1)j} + x_{i(j-1)} + x_{i(j+1)} $. If $b_{y}>b_{x}$, the deterministic dynamics predict that the producers will always go extinct. 

\begin{figure}[t]
\setlength{\abovecaptionskip}{-2pt plus 3pt minus 2pt}
\begin{center}
\includegraphics[width=0.5\textwidth]{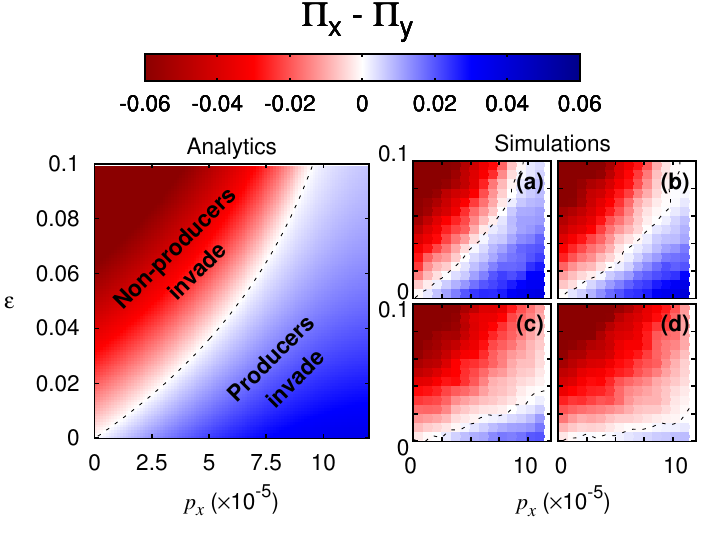}
\end{center}
\caption{Left panel: analytic results show that space amplifies stochastic selection reversal in the low dispersal, zero diffusion limit. The critical maximum cost $\varepsilon$ for selection reversal (see \eref{eq_eps_crit_space}) is plotted as black dashed line. Right panel: simulation results are shown for varied $m$ and $D$, averaged over $2000$ runs. Panel (a) gives results which in the low dispersal, zero diffusion limit ($m=3.7\times10^{-8}$, $D=0$), which match our theoretical predictions. Panel (b) gives the result with a set of biologically plausible parameters ($m=3.7\times10^{-7}$ and $D=2.2\times10^{-5}$ derived in \tref{tab:supmat_fig_parameter_list_1}). Panel (c) gives results in a system with high dispersal ($m=3.7\times10^{-5}$, $D=2.2\times10^{-5}$), while panel (d) shows the case of a system with high diffusion ($m=3.7\times10^{-7}$ and $D=2.2\times10^{-3}$). The number of patches is given by $C=16$ and the remaining parameters are listed in \tref{tab:supmat_fig_parameter_list_1}.}\label{fig_trade_off_16_islands}
\end{figure}

First we will discuss some important limit case behavior for this system. In the limit of large dispersal rate $m$ and diffusion rate $D$, the stochastic system behaves like a well-mixed population with a spatial scale $cR^{2}$ (i.e. the spatial structure is lost). In this case, as the size of the spatial system is increased, the effective population size also increases, and as a consequence selection reversal for producing phenotypes becomes less likely (see \eref{eq_eps_condition}).

We next consider the low-dispersal, zero diffusion limit. For sufficiently low dispersal, any incoming mutant will first either fixate or go to extinction locally before any further dispersal event occurs. Since each dispersal/invasion/extinction event resolves quickly, at the population level, the system behaves like a Moran process on a graph~\cite{nowak_2006}, with each node representing a patch. The `fitness' of a patch is the probability that it produces a migrant, and that that migrant successfully invades a homogeneous patch of the opposite type, following the approach used in \cite{houch_2012}. Denoting the `fitness' of producing and non-producing patches by $W_{x}$ and $W_{y}$ respectively, we have
\begin{equation}
  W_{x} = m N_{x} \phi_{x} \,, \qquad  W_{y} = m N_{y} \phi_{y}  \,, \label{eq_patch_fitness}
\end{equation}
where $N_{i}$ ($i = x,y$) is the mean carrying capacity of phenotype $i$ in a homogeneous patch, and $\phi_{i}$ are the invasion probabilities of a type $i$ mutant in a type $j\neq i$ patch. The fixation probabilities of a homogeneous patch in a population of the opposite phenotype can now be calculated using standard results~\cite{nowak_2006}. Let $\rho_{i}$ ($i = x,y$) denote the probability that type $i$ takes over the metapopulation when starting from one patch of type $i$ in a population otherwise comprised entirely of patches of the opposite phenotype. Then
\begin{equation}
 \rho_{i} = \frac{1 - r_{i}^{-1} }{ 1 - r_{i}^{-C} } \,, \quad \textrm{for} \quad i = x,y \quad \textrm{and} \quad r_{x} = \frac{W_{x}}{W_{y}} \,, \quad r_{y} = \frac{W_{y}}{W_{x}} \,. \label{eq_regular_graph_fix}
\end{equation}
If we start from a single invading mutant, the probability that it takes over the entire population (i.e. invasion probability) is the product between the probability that it takes over its home patch $\phi_{i}$, and the probability that the newly invaded home patch fixates into the metapopulation, $\rho_{i}$:
\begin{eqnarray}
 \Pi_{x} = \phi_{x} \rho_{x} \,, \quad  \Pi_{y} = \phi_{y} \rho_{y} \,. \label{eq_pi}
\end{eqnarray}
In the infinite patch limit ($C\rightarrow\infty$), $\rho_{x}$ and $\rho_{y}$ depend on $r_{x}$, the patch fitness ratio defined in \eref{eq_regular_graph_fix}. If $r_{x}>1$, $\rho_{x}\rightarrow1-r_{x}^{-1}$ and $\rho_{y}\rightarrow0$, whereas if $r_{x}<1$ the converse is true. This means that, in the infinite patch, low dispersal, zero diffusion limit, the condition for the stochastic reversal of selection is weakened from $\phi_{x} > \phi_{y}$ to
\begin{eqnarray}
 N_{x} \phi_{x} > N_{y} \phi_{y} \,. \label{eq_weakening_threshold}
\end{eqnarray}
Spatial structure therefore has the ability to enhance the stochastic reversal observed in the small well-mixed system. An approximate analytic form for the above condition can be obtained in terms of the original parameters;
\begin{equation}
 \varepsilon < 2 \frac{ \kappa }{ b R^{2} }  \log \left[  \frac{\delta  \kappa }{ \delta \kappa - p_{x} r }  \right]  \,. \label{eq_eps_crit_space}
\end{equation}

Once again, our analytical results are well supported by simulations (see \fref{fig_trade_off_16_islands}). The critical production rate for the invasion probability of producers to exceed that of non-producers has been decreased, as predicted by Eqs.~(\ref{eq_eps_condition}) and (\ref{eq_eps_crit_space}). Producers can therefore withstand higher production costs in spatially structured environments.

\begin{figure}[t]
\setlength{\abovecaptionskip}{-2pt plus 3pt minus 2pt}
\begin{center}
\includegraphics[width=0.5\textwidth]{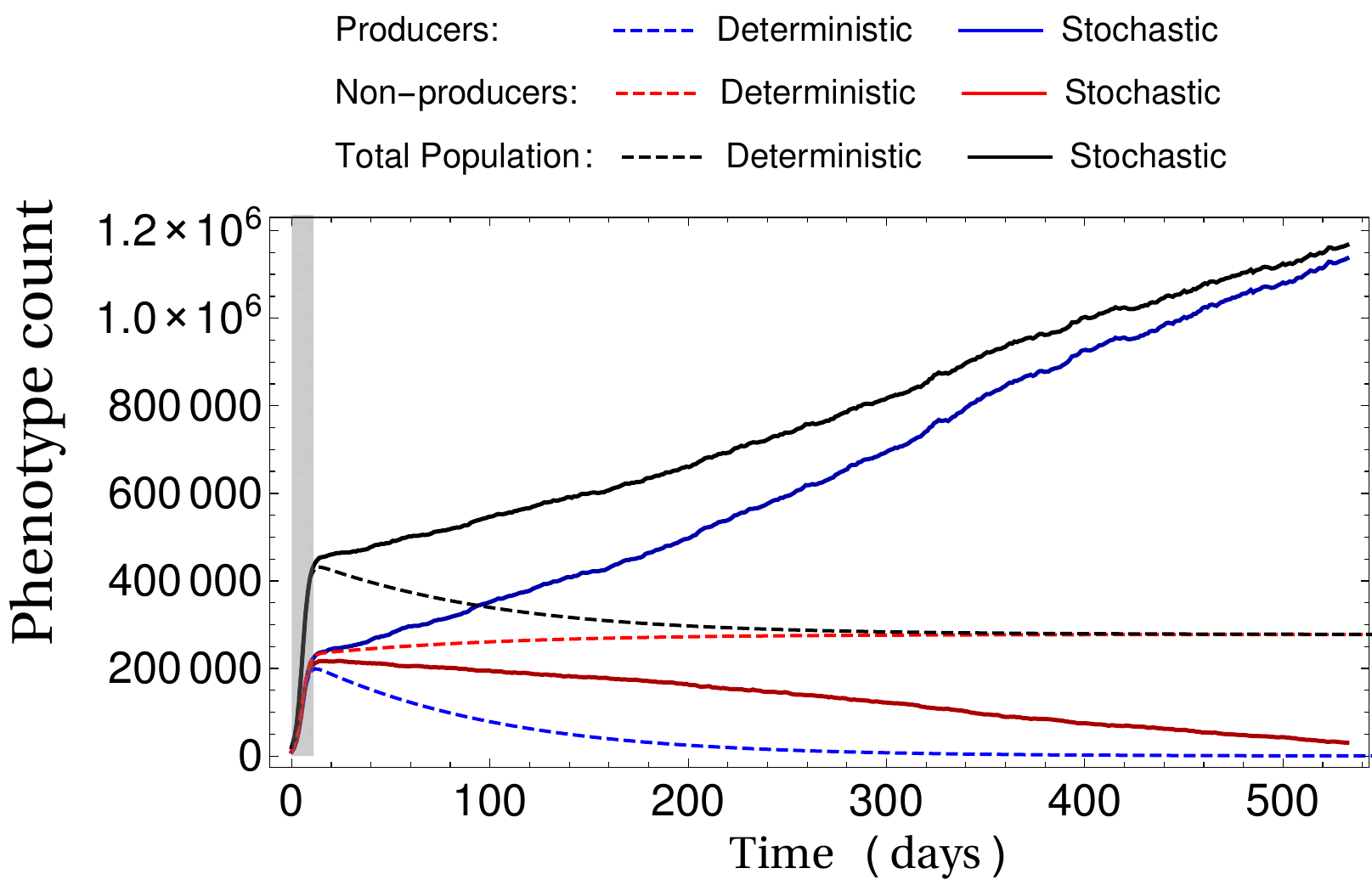}
\end{center}
\caption{Demographic stochasticity at the local `patch' scale profoundly alters the system dynamics at the population level. Results are obtained from stochastic and deterministic ($R\rightarrow \infty $) simulations of \eref{eq_spatial_ODEs} with a grid of $C=100\times100$ patches, $p_{x}=1\times10^{-4}$, $\varepsilon=0.02$, $m=3.7\times10^{-5}$ and the remaining parameters taken from \tref{tab:supmat_fig_parameter_list_1}. Initial conditions are a single producer and non-producer on each patch. The initial (fast) transient collapse to the SM occurs occurs in the shaded gray region. Following this, the deterministic system slowly moves along the slow manifold until the non-producers fixate, whereas in the stochastic system, the producers experience a selective pressure in their favor. For dynamics at the patch level, see Supplementary Information movie S1.}\label{fig_10000_islands}
\end{figure}

It is important to note that while \eref{eq_weakening_threshold} depends on the mean number of producers and non-producers on a homogeneous patch ($N_x$ and $N_y$), it is independent of the number of individuals in the entire metapopulation in the large $C$ limit. The interaction between these two spatial scales leads to results that can appear counter-intuitive. Demographic noise, as we have discussed, leads to producing patches being `more fit' at the patch level (see \eref{eq_patch_fitness}). However, when a large number of patches is considered, the demographic noise at the metapopulation level is reduced. This leads to the system following trajectories that appear deterministic at the level of the metapopulation, even though the path they follow is entirely the result of demographic stochasticity at the within-patch level (see \fref{fig_10000_islands}). The movie S1 (see \Aref{sec:supmat_sec_movies}) displays the individual dynamics of the patches that comprise the trajectory illustrated in \fref{fig_10000_islands}.

Away from the small dispersal, zero diffusion limit, the dramatic selection reversal predicted by the analytical results is clearly weakened (see \fref{fig_trade_off_16_islands}). Though selection reversal is still found across a range of $m$ and $D$ values, if either dispersal or diffusion are too high, the selection reversal breaks down. It is therefore important to understand what order of magnitude estimates for the values of $m$ and $D$ may be biologically reasonable.

\subsection{Insights from \fontsize{10}{11}{\textit{S.\,cerevisiae}}} In the following section, we will attempt to contextualize our model with reference to a \textit{S. cerevisiae} yeast system, which has been previously identified as a biological example of a population that features public good producers and non-producers. The model we have presented is general and therefore it could not capture the full biological detail of this particular system. For instance, it has been noted that some degree of privatization of the public good occurs in even the well-mixed experimental system~\cite{gore_2009}, a behavior we do not consider in our model. However, setting our model in this context can provide some insight into the scenarios in which we might expect stochastic selection reversal to be a biologically relevant phenomenon.

An \textit{S.\,cerevisiae} yeast cell metabolizes simple sugars, such as glucose, in order to function. However, when simple sugars are scarce, the yeast can produce invertase, an enzyme that breaks down complex sugars, such as sucrose, to release glucose~\cite{maclean_2010}. Invertase is produced at a metabolic cost and, since digestion of sucrose occurs extracellularly, most of the benefits of its production are shared by the population. Specifically in the case of S.~cerevisiae, $SUC2$, the wild-type strain, produces invertase, while the lab cultured mutant $suc2$ does not~\cite{maclean_2008}. In terms of our model parameters, the baseline birth rates, $b_{x}$ and $b_{y}$, represent respectively $SUC2$ and $suc2$ reproduction in the absence of invertase. This could be understood as arising from yeast directly metabolizing sucrose (a less energetically beneficial metabolic route~\cite{maclean_2010}) or as the result of some extrinsically imposed low glucose concentration in the system. The rate $r$ would then represent the additional birth rate in the presence of invertase. The form of our specified reactions (see \eref{eq_reactions_1}), assumes that the presence of invertase leads directly to a yeast reproduction event. In reality invertase must break down the sucrose into glucose, and then slowly absorb the glucose. We are therefore essentially assuming that the sucrose is abundant, its breakdown by invertase instantaneous, and the glucose absorption rapid and occurring in discrete packets, with each packet absorbed leading to a reproduction event.

In the well-mixed system, our analytic predictions indicate that stochastic selection reversal can occur only if the population is very small. Since this is an unrealistic assumption in the case of yeast cultures, we would predict that non-producers should come to dominate a well-mixed population. In a spatially structured population however, this constraint is relaxed since it only requires small interaction regions. For \textit{S.\,cerevisiae}, we can obtain order of magnitude estimates for the majority of parameters in our model, including public good diffusion (see \Aref{sec:supmat_sec_params}). Using these estimates together with our analytic results for the spatial public goods system, we find that stochastic selection reversal could feasibly be an important phenomenon for promoting the evolution of microbial public goods production in spatial settings (see \fref{fig_trade_off_16_islands}, Panel b). Given this, we now consider the results of a spatial experiment on \textit{S. cerevisiae}, and ask how its results might be interpreted in light of the insights developed with our simple model.

\begin{figure}[t]
\setlength{\abovecaptionskip}{-2pt plus 3pt minus 2pt}
\begin{center}
\includegraphics[width=0.5\textwidth]{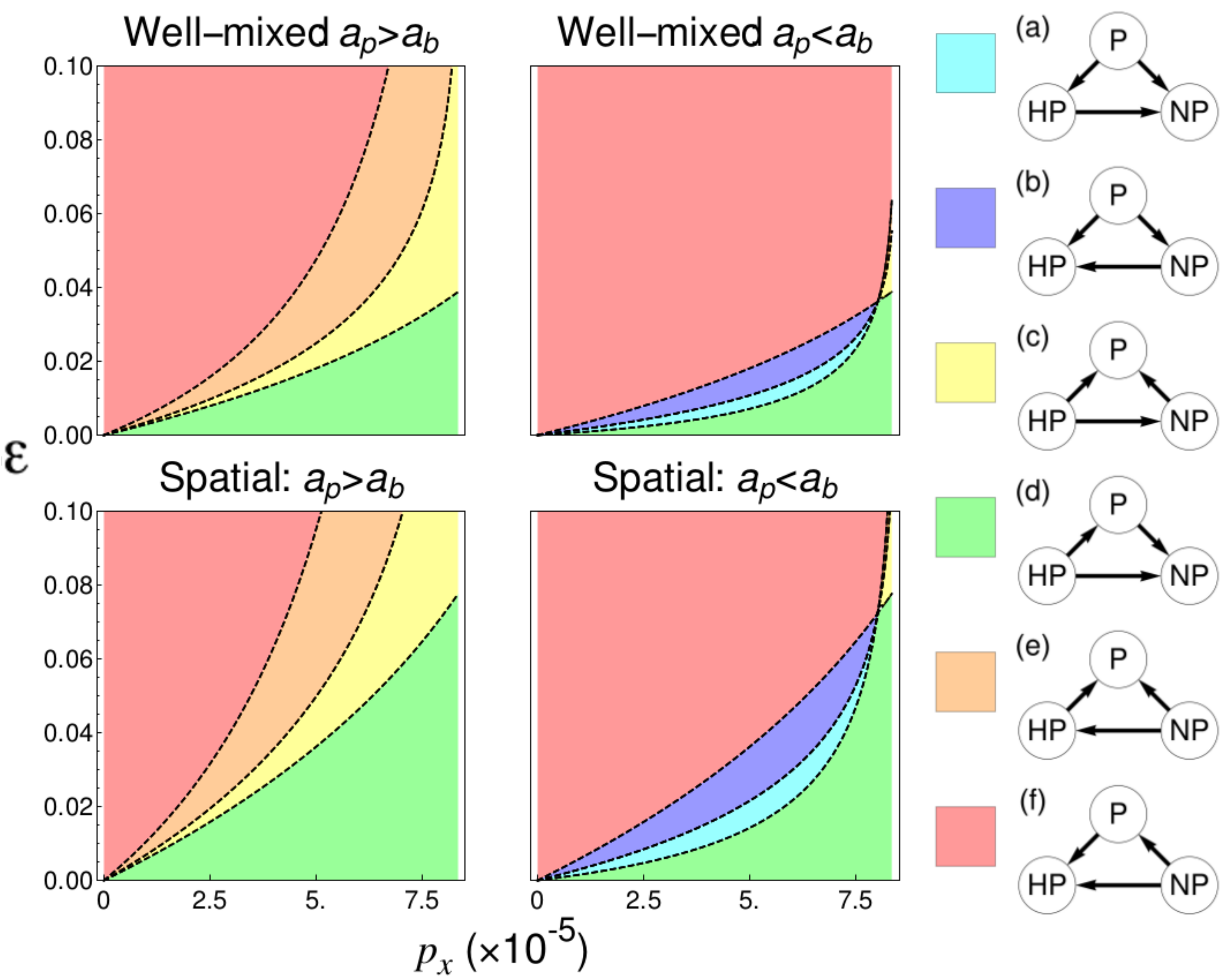}
\end{center}
\caption{Plots of the pairwise invariability scenarios possible for non-producing, producing and hyper-producing phenotypes. Arrows point away
from the dominant phenotype in a pair, which is defined as that with a larger invasion probability (Fig. S3). Non-transitive dynamics are not possible. It is possible however for an optimal intermediate good production rate to emerge (cyan and blue regions), if $a_{p}<a_{b}$. In this scenario the hyper-producer receives diminishing good production as a function of cost to birth rate compared to the producer. Left panels, $a_{b}=1.3$ and $a_{p}=1.5$. Right panels, $a_{b}=3$ and $a_{p}=1.5$. Remaining parameters given in \tref{tab:supmat_fig_parameter_list_1}. }\label{fig_pairwise_invade}
\end{figure}

In \cite{maclean_2008}, $SUC2$ and $suc2$ were experimentally competed on an agar plate. It was found that non-producing $suc2$ could not invade from rare ($1\%$ of initial yeast population), and in fact decreased in frequency, becoming undetectable at long times (around 800 generations). This suggests that in a spatial setting, invertase producing $SUC2$ yeast are robust to invasions, which is in qualitative agreement with our theoretical predictions. The experiments yielded an additional result, the appearance of a hyper-producing mutant. This hyper-producing phenotype produced invertase at approximately $1.5$ times the rate of standard producers and existed at higher densities. The hyper-producer appeared to evolve naturally and establish robust colonies during the competition experiments between non-producers and producers. However, when separate competition experiments were conducted between the hyper-producers and the producers, the hyper-producers failed to demonstrate any appreciable fitness advantage over the producers. This potentially suggests an optimal invertase production rate, whereby the hyper-producers managed to establish and grow during the $SUC2$-$suc2$ competition experiments by exploiting non-producing regions due to a relative fitness advantage, but could not invade regions of space occupied by producers. Interestingly, our model also predicts that an intermediate optimal production rate may exist, depending on how the cost of production scales with the production rate. Suppose a hyper-producer, $U$, produces at a rate $p_{u}=a_{p}p_{x}$, paying a metabolic cost $a_{b}\varepsilon$ to its birth rate, such that $b_{u}=b(1-a_{b}\varepsilon)$. The pairwise invasion probabilities of each phenotype can then be calculated (see Supplementary Information, \ignore{\sref{Ssec:supMat_sec_pairwise}}Section\,S.4). We define the fitter phenotype in a pair as that with the larger invasion probability. The potential fitness rankings are investigated in \fref{fig_pairwise_invade} as a function of $p_{x}$ and $\varepsilon$, (which we recall also alter $p_{u}$ and $b_{u}$). We draw particular attention to the right panels, in which $a_{b}>a_{p}$. In this scenario, the hyper-producers pay a disproportionate cost for their increased production rate compared to the producers. This can be interpreted as diminishing returns for production. In this case, there exist regions where the producer is the optimal phenotype (regions (a) and (b), in blue and cyan respectively). Specifically, scenario (a) displays a similar behavior to that observed in \cite{maclean_2008}, in which producers win out over both non-producers and hyper-producers, but hyper-producers are more likely to invade non-producing populations.

\section{Generality of results}

We have shown that demographic stochasticity can reverse the direction of selection in a public good model. In this section we will show that the mechanism responsible for this phenomenon is by no means particular to this model. We consider a general scenario, with a phenotype $X_{1}$, which is at the focus of our study, and a number of discrete ecosystem constituents, $E_{i}$. In the public good model for instance, we would label the public good itself as an ecosystem constituent, however more generally this could be a food source, a predator or anything else that interacts with the phenotypes. The state of the ecosystem influences the birth and death of the phenotype and in turn the presence of the phenotype influences the state of the ecosystem, altering the abundances of the constituents. We assume that the system lies at a unique, stable stationary state, precluding the possibility of periodic behavior. Suppose that a new phenotype, $X_{2}$, arises. We assume that the second phenotype is only slightly better at exploiting the ecosystem than $X_{1}$, though its influence on the ecosystem may be very different. For instance, in the public goods model, non-producers have a small birth-rate advantage over producers, but do not produce the public good. Which phenotype is more likely to invade and fixate in a resident population of the opposite type?

The stochastic model for this system can be constructed in a similar manner to the public good model; the dynamics are described by a set of probability transition rates (analogous to \eref{eq_reactions_1}). We restrict the transitions by specifying that although the two phenotypes compete, there is no reaction that instantaneously changes both of their numbers in the population. This final condition simply means that they should not, for instance, be able to mutate from one type to another during their lifetime, or to prey on each other. A parameter $R$ is introduced, to once again govern the typical scale of the system. The model is analyzed in the mesoscopic limit, by introducing $(x_{1},x_{2},\bm{e}) = (n_{x1},n_{x2},\bm{n}_{e})/R^{2}$ and applying the diffusion approximation. For large but finite $R$, the mesoscopic description takes the form
\begin{eqnarray}
 \dot{ x_{1} } &=& x_{1} F^{(0)}(\bm{x},\bm{e}) - \varepsilon x_{1} F^{(\varepsilon)}(\bm{x},\bm{e}) + R^{-1}\eta_{1}(t) \, , \nonumber \\
 \dot{ x_{2} } &=& x_{2} F^{(0)}(\bm{x},\bm{e}) + R^{-1} \eta_{2}(t) \, , \label{eq_general_SDEs} \\
 \dot{ e_{i} } &=& F_{i}(\bm{x},\bm{e}) + R^{-1} \beta_{i}(t) \,, \quad \forall \, i = 3 , \ldots J \,, \nonumber
\end{eqnarray}
where $\varepsilon$ is small and governs selective pressure against $X_{1}$. The assumption that there is no reaction that instantaneously changes the number of both phenotypes ensures that the correlation structure of the noise terms takes the form
\begin{eqnarray}
& \langle \eta_{1}(t) \eta_{1}(t') \rangle = \delta(t-t') x_{1} H^{(0)}(\bm{x},\bm{e}) \,,& \nonumber \\
& \langle \eta_{2}(t) \eta_{2}(t') \rangle = \delta(t-t') x_{2} H^{(0)}(\bm{x},\bm{e}) \,,& \quad \langle \eta_{1}(t) \eta_{2}(t') \rangle = 0 \nonumber \,,
\end{eqnarray}
with $\varepsilon$ taken to be of order $R^{-2}$. This assumption, made here to isolate the effect of varying carrying capacity from any other intraspecies dynamics, means that while the magnitude of fluctuations in the number of both phenotypes is dependent on the state of the system, $(\bm{x},\bm{e})$, the fluctuations themselves are not correlated with each other. Restrictions on the microscopic model that yield the above SDE description are addressed more thoroughly in \Aref{sec:supmat_sec_general_results}. The form of \eref{eq_general_SDEs} makes the nature of the system we describe more clear; it consists of two competing phenotypes, which reproduce according to replicator dynamics~\cite{hofbauer_1998} with equal fitness at leading order in $\varepsilon$. 

In the special case $\varepsilon=0$, both phenotypes are equally fit, regardless of their influence on the ecosystem variables $e_{i}$. The degeneracy of the dynamics in $x_{1}$ and $x_{2}$ ensures the existence of a deterministic CM. We assume that the structure of $F^{(0)}(\bm{x})$ and $F_{i}(\bm{x})$ is such that the CM is one-dimensional (there are no further degenerate ecosystem variables) and that it is the only stable state in the interior region $x_{i}>0$. A separation of timescales is present if the system collapses to the CM much faster than the stochastic dynamics. In practical terms, the timescale of collapse can be inferred as the inverse of the non-zero eigenvalues of the system, linearised about the CM~\cite{constable_2013}, while the timescale of fluctuations  will be of order $R^{-2}$~\cite{constable_2014_phys}. When $\varepsilon>0$, the timescale elimination procedure can still be applied if $\varepsilon\approx\mathcal{O}\left( R^{-2} \right)$ . The effective one-dimensional description of the system now takes the form 
\begin{equation}
 \dot{x_{1}} =  - \varepsilon \, \mathcal{D}(x_{1}) + R^{-2} \,\mathcal{S}(x_{1}) + R^{-1} \zeta(t) \,, \label{eq_general_reduced_PG_SDE}
\end{equation}
where the term $ \mathcal{D}(x_{1})$ is the deterministic contribution to the effective dynamics and $\mathcal{S}(x_{1})$ is the stochastic contribution, while $\zeta(t)$ is an effective noise term. The form these functions take is dependent on $F^{(0)}(\bm{x},\bm{e})$, $F^{(\varepsilon)}(\bm{x},\bm{e})$ and $F_{i}(\bm{x},\bm{e})$, as well as the noise correlation structure, $H^{(0)}(\bm{x},\bm{e})$; however it is independent of the structure of the demographic noise acting on the ecosystem variables (see Eqs.\,(\ref{eq:supmat_eq_general_Abar_det}), (\ref{eq:supmat_eq_Abar_stoch_2}) and (\ref{eq:supmat_eq_general_Bbar})). 

The core assumption we have made to derive \eref{eq_general_reduced_PG_SDE} is essentially that the system's ecological processes act on a faster timescale than its evolutionary processes. Even in this general setting, insights about the system's stochastic dynamics can still be drawn (see \Aref{sec:supmat_sec_general_results}). If $\varepsilon=0$, the fixation probability of phenotype $X_{1}$ is independent of the initial conditions of the ecosystem variables $\bm{e}$. In fact it is equal to the initial fraction of $X_{1}$ in the population, $ n_{10} / (n_{10} + n_{20} )$. The invasion probability of mutant $X_{1}$ phenotype fixating in a resident $X_{2}$ population however depends on the stationary state of the $X_{2}$ population; this defines the initial invasion conditions (the denominator for the fixation probability of $X_{1}$). Denoting by $N_{1}$ and $N_{2}$ the average numbers of phenotypes $X_{1}$ and $X_{2}$ in their respective stationary states, we find $\phi_{1} = 1/N_{2}$ and $\phi_{2} = 1/N_{1}$, generalizing \eref{eq_phi_quasineutral}. Therefore, for $\varepsilon=0$ the phenotype that exists at higher densities is more likely to invade and fixate than its competitor, a consequence of its robustness to invasions. This result holds for any choice of finite $R$. In an ensemble of disconnected populations subject to repeated invasions, we would observe the emergence of high density phenotypes if this phenotype does not carry a cost. While this seems like a reasonable and indeed natural conclusion, it is one entirely absent from the deterministic analysis. 

If $\varepsilon>0$, general results for the phenotype fixation probabilities cannot be obtained. However, if $N_{1}>N_{2}$, in the limit $\varepsilon\rightarrow0$ we have shown that $\phi_{1}>\phi_{2}$. From this, it can be inferred that the term $\mathcal{S}(x_{1})$ is positive on average along the slow manifold (see \eref{eq:supmat_integral_S}). Therefore, if phenotype $X_{1}$ exists at higher densities in isolation than phenotype $X_{2}$, there will exist a stochastically-induced pressure favoring the invasion of phenotype $X_{1}$. Meanwhile, by construction we expect the form of $\mathcal{D}(x_{1})$ to be positive, since phenotype $X_{1}$ exploits the ecosystem environment less effectively than phenotype $X_{2}$. There is therefore a trade-off for competing phenotypes between increasing their phenotype population density  and increasing their per capita growth rate. Note that the noise-induced selection function $\mathcal{S}(x_{1})$ need not be strictly positive; indeed it may become negative along regions of the SM. This potentially allows for stochastically induced `fixed points' along the SM, around which the system might remain for unusually large periods of time. This may provide a theoretical understanding of the coexistence behavior observed in~\cite{behar_2015}.

The term $\mathcal{S}(x_{1})$ is moderated by factor $R^{-2}$ (see \eref{eq_reduced_PG_SDE}), or more physically, the typical size of the population. The stochastically induced selection for the high-density phenotype therefore becomes weaker as typical system sizes increase. The trade-off will be most crucial in small populations, or as illustrated in the public good model, systems with a spatial component. If the phenotypes and ecosystem variables move sufficiently slowly in space, the results of Eqs.~(\ref{eq_pi}) and (\ref{eq_weakening_threshold}) can be imported, with the understanding that $\phi_{1}$ and $\phi_{2}$ must be calculated for the new model under consideration. 

It is worth noting that the precise functional form of $\phi_{1}$ and $\phi_{2}$ identified in the deterministically neutral case ($\varepsilon=0$) is dependent on the assumption that phenotype noise fluctuations are uncorrelated. While correlated fluctuations (for instance resulting from mutual predation of the phenotypes) can still be addressed with similar methods to those employed here, there is then the potential for the emergence of further noise-induced selection terms (see \Aref{sec:supmat_sec_comp_models}). Careful specification of the phenotype interaction terms is therefore needed to determine to what degree these additional processes might amplify or dampen the induced selection we have identified.

\section{Discussion}

In this paper, we have shown that stochastic effects can profoundly alter the dynamics of systems of phenotypes that change the carrying capacity of the total population. Most strikingly, selection can act in the opposite direction from that of the deterministic prediction if the phenotype that is deterministically selected for also reduces the carrying capacity of the population. The methods used to analyze the models outlined in the paper are based on the removal of fast degrees of freedom~\cite{parsons_rogers_2015}. The conclusions drawn are therefore expected to remain valid as long as the rate of change of the phenotype population composition occurs on a shorter timescale than the remaining ecological processes. 

By illustrating this phenomenon in the context of public good production, we have revealed a mechanism by which the dilemma of cooperation can be averted in a very natural way: by removing the unrealistic assumptions of fixed population size inherent in Moran-type game theoretic models. The potential for such behavior has been previously illustrated with the aid of a modified Moran model~\cite{houch_2012} and a single variable Wright-Fisher type model~\cite{houch_2014} that assumes discrete generations. However we have shown that the mechanism can manifest more generally in multivariate continuous time systems. Our analysis may also provide a mathematical insight into the related phenomenon of fluctuation-induced coexistence that has been observed in simulations of a similar public good model featuring exogenous additive noise~\cite{behar_2015}: such coexistence may rely on a similar conflict between noise-induced selection for producing phenotypes and deterministic selection against them.

For biologically reasonable public good production costs, selection reversal is only observed in systems that consist of a very small number of individuals. However, by building a metapopulation analogue of the model to account for spatial structure, the range of parameters over which selection reversal is observed can be dramatically increased, so long as public good diffusion and phenotype dispersal between populations are not large. Two distinct mechanisms are responsible for these results. First, including spatial structure allows for small, local effective population sizes, even as the total size of the population increases. This facilitates the stochastic effects that lead to selection reversal. Second, since producer populations tend to exist at greater numbers (or higher local densities) they produce more migrants. The stochastic advantage received by producers is thus amplified, as not only are they more robust stochastically to invasions, but also more likely to produce invaders. Away from the low-dispersal, zero public good diffusion limit, the effect of selection reversal is diminished, but is still present across a range of biologically reasonable parameters. The analytical framework we have outlined may prove insightful for understanding the simulation results observed in \cite{behar_2014}, where a similar metapopulation public good model was considered. In addition to fixation of producers (in the low dispersal-diffusion limit) and fixation of non-producers (in the high dispersal-diffusion limit), \cite{behar_2014} observed an intermediate parameter range in which noise induced coexistence was possible. Though our model does not feature such a regime, extending our mathematical analysis to their model would be an interesting area for future investigations. However it must be noted that coexistence in a stochastic setting is inherently difficult to quantify analytically, as for infinite times some phenotype will always go extinct.

That space can aid the maintenance of cooperation is well known~\cite{tarnita_review_2009,wakano_2011}. Generally, however, this is a result of spatial correlations between related phenotypes, so that cooperators are likely to be born neighboring other cooperators (and share the benefits of cooperation) while defectors can only extract benefits at the perimeter of a cooperating cluster. This is not what occurs in the model presented in this paper. Indeed, while we have assumed in our analytic derivation of the invasion probability that dispersal is small enough that each patch essentially contains a single phenotype, we find that the phenomenon of selection reversal manifests outside this limit (see \Aref{sec:supmat_sec_movies}, movie S2 in which a majority of patches contain a mix of producers and non-producers). Instead, producing phenotypes have a selective advantage due to the correlation between the fraction of producers on a patch and the total number of individuals on a patch, which provides both resistance to invasions and an increased dispersal rate. 

Most commonly in spatial game theoretic models of cooperation-defection, individuals are placed at discrete locations on a graph~\cite{nowak_1992,allen_2013}. In contrast, by using a metapopulation modelling framework we have been able to capture the effect of local variations in phenotype densities across space, which is the driver of selection amplification in our model. Nevertheless, the question that remains is which modelling methodology is more biologically reasonable. This clearly depends on the biological situation. However, in terms of test-ability, our model makes certain distinct predictions. In~\cite{allen_2013}, producers and non-producers were modeled as residing on nodes of a spatial network, with a public good diffusing between them. The investigation concludes that both lower public good diffusion and lower spatial dimensions (e.g. systems on a surface rather than in a volume) should encourage public good production, essentially by limiting the `surface area' of producing clusters. While our investigation certainly predicts that lower public good diffusion is preferable, stochastic selection reversal does not require that the spatial dimension of the system is low. In fact the result utilized in \eref{eq_regular_graph_fix} holds for patches arranged on any regular graph (where each vertex has the same number of neighbors), and thus could be used to describe patches arranged on a cubic, or even hexagonal, lattice.

In our final investigation, we have shown that stochastic selection reversal is not an artifact of a specific model choice, but may be expected across a wide range of models. These models consist of two phenotypes, competing under weak deterministic selection strength, reproducing according to replicator dynamics and interacting with their environment. Thus the phenomenon of selection reversal is very general; however, it depends strongly on how one specifies a selective gradient. We take one phenotype to have a stochastic selective advantage over the other if a single mutant is more likely to invade a resident population of the opposite type. Such a definition is also used in standard stochastic game theoretic models~\cite{nowak_2006}. A key difference here however (where the population size is not fixed) is that the invasion probability is not specified by a unique initial condition; we must also specify the size of the resident population. We have assumed that the invading mutant encounters a resident population in its stationary state. This is by no means an unusual assumption; it is the natural analogue of the initial conditions in a fixed population size model. Essentially it assumes a very large time between invasion or mutation events, an approach often taken in adaptive dynamics~\cite{waxman_2005}. 

If instead we assumed a well-mixed system far from the steady state, our results would differ. For instance, suppose the system initially contains equal numbers of the two phenotypes. For the case when the two phenotypes have equal reproductive rates ($\varepsilon=0$), the phenotypes have equal fixation probability. For $\varepsilon>0$, the phenotype with the higher birth rate has the larger fixation probability, regardless of its influence on the system's carrying capacity. This apparent contradiction with the results we developed in the body of the paper echos the observations of $r-K$ selection theory~\cite{pianka_1970}: selection for higher birth rates ($r$-selection) acts on frequently disturbed systems that lie far from equilibrium, while selection for improved competitive interactions or carrying capacities ($K$-selection) acts on rarely disturbed systems. In addition, $r-K$ selection theory suggests that $K$-selected species are typically larger in size and, as a consequence, consist of a lower number of individuals~\cite{reznick_2002}. This indicates a further parallel with our stochastic model framework, since selection for higher carrying capacities requires that the typical number of individuals (of both the low and high carrying capacity phenotypes) is small. Though the mechanism that leads us to these conclusions is distinct, our stochastic analysis provides a complementary view of $r-K$-selection theory, which may be applicable to simple microorganisms. In exploring this analogous behavior further, future investigations may also benefit from considering the results of~\cite{parsons_quince_2010}, where it was shown that stochastically induced selection can change direction near carrying capacity.

Although we have implicitly developed our results in the low mutation limit, including mutation explicitly in the modeling framework is possible. This would be an interesting extension to the framework. In the well-mixed scenario, it is likely that the inclusion of mutation will complicate the intuition developed here: while larger populations are more robust to invasions, they are also more prone to mutations, by virtue of their size. While this may be offset by the additional benefits garnered in the spatial analogue of the model, a complex set of timescale-dependent behaviors is likely to emerge.

Finally, we propose a rigorous analytical investigation of existing models that conform to the framework we have outlined; an example is the work conducted in \cite{behar_2014,behar_2015}, which we believe to be mathematically explainable within our formalism. In the context of induced selection, whereby deterministically neutral systems become non-neutral in the stochastic setting, similar ideas have already been extended to disease dynamics \cite{kogan_2014} and the evolution of dispersal \cite{lin_mig_1,lin_mig_2}. The extension of selection reversal to such novel ecological models may provide further insight. Furthermore, this general scheme may be of relevance to many other systems in ecological and biological modeling, such as cancer, for which the evolution of phenotypes that profoundly alter cell carrying capacity can be of primary importance.

\begin{acknowledgments}
TR acknowledges funding from the Royal Society of London.
\end{acknowledgments}

\appendix

\section{Obtaining the SDE system from the microscopic individual based model}\label{sec:supmat_sec_obtain_SDE_system}

We begin with a model consisting of a discrete number of entities, two phenotypes of a species, $X$ and $Y$ and a public good $Q$. They interact according to the transitions
\begin{eqnarray}
& X \overset{b_{x}}{\underset{\kappa/R^{2}}{ \myrightleftarrows{\rule{0.4cm}{0cm}} } } X+X\,,\quad Y + X \xrightarrow{\kappa/R^{2}} X \,, & \nonumber \\
& Y \overset{b_{y}}{\underset{\kappa/R^{2}}{ \myrightleftarrows{\rule{0.4cm}{0cm}} } }  Y+Y\,,\quad Y + X \xrightarrow{\kappa/R^{2}} Y \,,  & \,, \nonumber \\  
& X+Q\xrightarrow{r/R^{2} } X+X+Q\,,\quad   Y+Q\xrightarrow{r/R^{2} } Y+Y+Q&\,,\nonumber \\
& X\xrightarrow{p_{x}}X+Q\,, \quad Q\xrightarrow{\delta}\varnothing \,. &\label{supmat_eq_reactions}
\end{eqnarray}
The term $R^{-2}$ occurs in all terms involving two reactants. It thus controls the interaction probability between instances of the phenotypes and the public good. Taking larger $R$ decreases the interaction probability of phenotypes $X$ and $Y$ and the public good and allowing the populations to grow to greater numerical abundances. The parameter $R$ can thus be understood as a measure of the spatial scale of the system; when $R$ is increases, the probability of interactions in the well-mixed system is decreased while the number of individuals the system can contain is increased.

Let us denote $\bm{n}=(n_{x}$, $n_{y}$, $n_{q})$ the numbers of $X$, $Y$ and $Q$ respectively. Then the dynamics of this system can be described by the set of partial difference equations
\begin{equation}
 \frac{d P(\bm{n},t)}{d t} =  \sum_{\bm{n}'\neq \bm{n}} \left[  T(\bm{n}|\bm{n}')P(\bm{n}',t) - T(\bm{n}'|\bm{n})P(\bm{n},t) \right] \, ,\label{eq:supmat_meqn_general}
\end{equation}
where $P(\bm{n},t)$ is the probability of the state being in state $\bm{n}$ at time $t$, and $T(\bm{n}'|\bm{n})$, the probability transition rate, is the probability per unit time of transitioning from state $\bm{n}$ to $\bm{n}'$. Formally this is known as the master equation~\cite{van_kampen_2007}. Given the reactions \eref{supmat_eq_reactions} the probability transition rates can be expressed as 
\begin{eqnarray}
&T_1(n_x + 1,n_y,n_{q}|n_x,n_y,n_{q}) = b_{x}n_{x} + \dfrac{ r }{ R^{2} } n_{x} n_{q} \, ,& \nonumber \\
&T_2(n_x,n_y + 1,n_{q}|n_x,n_y,n_{q}) = b_{y}n_{y} + \frac{ r }{ R^{2} } n_{y} n_{q} \, ,&  \nonumber \\
&T_3(n_x - 1,n_y,n_{q}|n_x,n_y,n_{q}) = \dfrac{ \kappa }{ R^{2} } n_{x} \left( n_{x} + n_{y} \right) \,, & \nonumber \\
&T_4(n_x,n_y - 1,n_{q}|n_x,n_y,n_{q}) = \frac{ \kappa }{ R^{2} } n_{y} \left( n_{x} + n_{y} \right) \,,  & \nonumber \\
&T_5(n_x,n_y,n_q + 1|n_x,n_y,n_{q}) = p_{x} n_{x}  \,, & \nonumber \\
&T_6(n_x,n_y,n_{q}-1|n_x,n_y,n_{q}) =  \delta n_{q} \,.  \label{supmat_eq_trans}
\end{eqnarray}

Let us now make a change of variables into the scaled expressions $\bm{x}=(x,y,q)=(n_{x},n_{y},n_{q})/R^{2}$. Substituting the probability transition rates into \eref{eq:supmat_meqn_general}, we find recurrent factors of $1/R^{2}$ appearing in the resulting expression. These terms are associated with the local transitions from state $\bm{n}$ to the surrounding states. If $R^{2}$ is sufficiently large, the population grows larger (as the crowding terms in \eref{supmat_eq_reactions} grow small). We may then Taylor expand \eref{eq:supmat_meqn_general} in $R^{-1}$, assuming that the variables $(x,y,q)$ are approximately continuous~\cite{gardiner_2009}. Truncating at second order in $R^{-4}$, we arrive at a partial differential equation for $p(x,y,q,t)$ of the form
\begin{eqnarray}
\frac{\partial p(\bm{x},t)}{\partial t} = &-&  \frac{1}{R^{2}} \sum_{i} \frac{\partial}{\partial x_{i}} \left[A_{i}(\bm{x})p(\bm{x},t)\right]  \nonumber \\ &+& \frac{1}{ 2 R^{4} }\sum_{i,j} \frac{\partial^{2}}{\partial x_{i} \partial x_{j}} \left[B_{ij}(\bm{x})p(\bm{x},t)\right] \, , \label{eq:supmat_eq_FPE_general} \\
\bm{x} &=& (x_{1},x_{2},x_{3}) \equiv (x,y,q) \,.  \nonumber
\end{eqnarray}
This is a diffusion approximation in a population genetics context~\cite{crow_kimura_into}, but more generally is akin to the Kramers-Moyal expansion~\cite{gardiner_2009} or a nonlinear analogue of the van Kampen expansion~\cite{van_kampen_2007}. The forms of $\bm{A}(\bm{x})$ and $B(\bm{x})$, given transition rates \eref{supmat_eq_trans} are found to be~
\begin{eqnarray}
 A_{x}(\bm{x}) &=& x \left( b_{x} + r q  - \kappa ( x + y ) \right) \,, \nonumber \\
 A_{y}(\bm{x}) &=& y \left( b_{y} + r q -  \kappa ( x + y ) \right) \,, \nonumber \\
 A_{q}(\bm{x}) &=& p_{x} x  - \delta q \,, \label{supmat_eq_A_supMat}
\end{eqnarray}
and 
\begin{eqnarray}
&B_{xx}(\bm{x}) = x \left( b_{x} + r q  + \kappa x  + \kappa y \right) \,, &\nonumber \\
&B_{yy}(\bm{x}) = y \left( b_{y} + r q  + \kappa x  + \kappa y \right) \,, &\nonumber \\
&B_{qq}(\bm{x}) = p_{x} x  + \delta q \,,  &\nonumber \\
&B_{ij} = 0  \quad \forall \quad i \neq j \,. & \label{supmat_eq_B_supMat}
\end{eqnarray}
Further, it can be shown that the above PDE is equivalent to the set of It\={o} SDEs~\cite{risken_1989}
\begin{equation}
 \frac{ \mathrm{d} \bm{x} }{ \mathrm{d} \tau}  = \bm{A}(\bm{x}) + \frac{1}{R} \bm{\eta}(\tau) \,, \label{supmat_eq_SDE_very_general}
\end{equation}
where $ \tau = t R^{2}$ and $\bm{\eta}(t)$ are Gaussian white noise terms with zero mean and correlations 
\begin{equation}
 \langle \eta_{i}(\tau) \eta_{j}(\tau') \rangle = \delta( \tau - \tau' ) B_{ij}(\bm{x}) \,. \label{supmat_eq_correlations}
\end{equation}
Notice that the correlations are multiplicative and thus dependent on the state of the system.

\section{Obtaining one-dimensional effective public good model}\label{supmat_sec_public_good_reduction}

In this section we seek to identify and remove the fast-modes of the SDE system \eref{supmat_eq_SDE_very_general}, and thus obtain an effective one-dimensional description of the dynamics. We make use of methods of fast-mode elimination described in \cite{parsons_rogers_2015}. Firstly we note that the deterministic nullcline for $q$ is given by 
\begin{equation}
q = \frac{ p_{x} x  }{ \delta } \equiv Z_{q}(x,y) \,. \label{supmat_eq_q_cm}
\end{equation}
Therefore, if the production and decay of public good occur much faster than the processes associated with the phenotypes, we would expect the public good to quickly attain this value, after which its dynamics would be slaved to those of $x$ and $y$. Notice that deterministically, substituting \eref{supmat_eq_q_cm} into \eref{supmat_eq_A_supMat} recovers a Lotka-Volterra competition model for two competing species. 

To make further analytic progress, we begin by considering the quasi-neutral limit in which $b_{x}=b_{y} \equiv b$. Under these conditions, the deterministic system exhibits a center manifold (CM) given by \eref{supmat_eq_q_cm} and
\begin{equation}
y = \frac{ \left[ b \delta - (\delta \kappa - r p_{x} ) x  \right] }{ \delta \kappa}  \equiv Z_{y}(x) \,. \label{supmat_eq_y_cm}
\end{equation}
The CM is stable for $ \kappa \delta > r p_{x}$, and we assume that this condition holds throughout the paper. Calculating the intersection of the center manifold at the boundaries $y=0$ and $x=0$ allows us to determine the mean population size in the quasi-neutral ($\varepsilon=0$) limit when it consists of only producers and non-producers respectively;
\begin{eqnarray}
 N_{x}^{(0)} = R^{2} K_{x}^{(0)} \,, \quad  K_{x}^{(0)} = \left( \frac{  b \delta  }{ \delta \kappa - r p_{x} } \right) \,, \\
 N_{y}^{(0)} = R^{2} K_{y}^{(0)} \,, \quad  K_{y}^{(0)} = \left( \frac{  b \delta  }{ \delta \kappa - r p_{y} } \right) \,. \label{eq:supmat_eq_NX_NY_0}
\end{eqnarray}
These parameters will be useful in the following analysis. 

Deterministically, the system comes to rest on a point along the CM (defined by Eqs.~(\ref{supmat_eq_q_cm}) and(\ref{supmat_eq_y_cm})), which depends on the system's initial conditions. When stochasticity is included, the CM ceases to exist in any true sense. However, when the noise is small (already assumed in the derivation of SDEs~(\ref{supmat_eq_SDE_very_general})) we can say that far from the CM, we expect the dynamics to be dominated by the deterministic collapse to the CM, while in the vicinity of the CM, we expect noise to play a more important role, driving the slow change in population composition until one or other of the phenotypes fixates. We wish to exploit this timescale separation, and obtain an effective description of the dynamics in terms of a single variable. 

To begin, we note that the stochastic dynamics along the CM has two components. First, noise can move the system neutrally \emph{along} the CM. Second, noise can take the system \emph{off} the CM, at which point we expect the deterministic component of the dynamics to become more prevalent, driving the system back to the CM. In order to capture the effect of both of these processes on the effective dynamics along the CM, we implement a non-linear projection of the stochastic system to the CM. Essentially this assumes that fluctuations which take the system away from the manifold are instantaneously mapped along deterministic trajectories back to the CM. In order to formalize this, the mapping $z=f(x,y,q)$ is introduced, where $f(x,Z_{y}(x),Z_{q}(x))=x$; that is $z$ gives the position on the CM, parameterized by $x$, which intersects a deterministic trajectory beginning at $(x,y,q)$. The mapping can be determined analytically from the observation that the quantity $x/y$ in \eref{supmat_eq_SDE_very_general} is invariant in this quasi-neutral ($b_{x}=b_{y}$) scenario. Therefore
\begin{equation}
 \frac{z}{Z_{y}(z)} = \frac{x}{y} \,, \quad z = \frac{ b \delta x }{  (\delta \kappa - p r ) x + \delta \kappa y } \,. \label{supmat_eq_PG_define_z}
\end{equation}

The effective dynamics for $z$ can now be straightforwardly calculated by differentiating \eref{supmat_eq_PG_define_z} with respect to $t$. One must note however that since the original SDE system is defined in the It\={o} sense, the normal rules of calculus no longer apply. Applying It\={o}'s rules of calculus appropriately~\cite{van_kampen_2007,parsons_rogers_2015}, we find that the effective dynamics along the CM take the following form
\begin{equation}
\dot{z} =  \frac{1}{R^{2}}\mathcal{S}( z ) + \frac{1}{R} \zeta (t) \,, \label{supmat_eq_quasineutral_effective_SDE}
\end{equation}
where 
\begin{eqnarray}
  \mathcal{S}( z ) &=&  \frac{1}{2}  \left( \frac{\partial^{2} z }{\partial x^{2} } B_{xx}(\bm{x}) + \frac{\partial^{2} z }{\partial y^{2} } B_{yy}(\bm{x})  \right)|_{x=z,y=Z_{y}(z),q=Z_{q}(z)} \,,  \label{supmat_eq_PG_general_Abar} \\
                   &=& \frac{ 2 p_{x} r }{ \delta } z \left\{   1 + \frac{1}{ b^{2} \delta^{2}  } z \left[  b \delta \left( 2 p_{x} r - \delta \kappa \right) + p_{x} r \left(  p_{x} r - \delta \kappa \right) z \right] \right\} \,,  \nonumber \\
		   &=& 2 b \left( \frac{ K_{x}^{(0)} - K_{y}^{(0)} }{ (K_{x}^{(0)})^{3} (K_{y}^{(0)})^{2} } \right) z \left( K_{x}^{(0)} - z \right) \left[ K_{x}^{(0)} K_{y}^{(0)} \right. \nonumber \\ &+& \left.  \left( K_{x}^{(0)} - K_{y}^{(0)} \right) z \right] \,, \label{supmat_eq_PG_clean_Abar}
\end{eqnarray}
and 
\begin{eqnarray}
 \langle \zeta(t) \rangle &=& 0 \,, \quad \langle \zeta(t) \zeta(t') \rangle = \delta( t - t' ) \mathcal{B}(z) \,, \nonumber 
\end{eqnarray}
with
\begin{eqnarray}
 \mathcal{B}(z) &=&\left(\left[ \frac{\partial z }{\partial x } \right]^{2}B_{xx}(\bm{x}) + \left[ \frac{\partial z }{ \partial y }\right]^{2}B_{yy}(\bm{x}) \right)|_{x=z,y=Z_{y}(z),q=Z_{q}(z)} \,, \label{supmat_eq_PG_general_Bbar}\\ 
                &=& 2 z  \left\{ b + \frac{1}{ b^{2} \delta^{3} } z \left[ b^2 \delta^2 \left( 3 p_{x} r - \delta \kappa \right)
\right. \right. \nonumber \\ &+& \left. \left. b p_{x} r \delta \left( 3 p_{x} r - 2 \delta \kappa \right) z  
		 p_{x}^2 r^2 \left( p_{x} r - \delta \kappa \right)z^{2} \right] \phantom{\frac{1}{2}}  \right\} \,, \nonumber \\
		&=&  2 b \left( \frac{ 1 }{ (K_{x}^{(0)})^{3} (K_{y}^{(0)})^{2} } \right) z \left( K_{x}^{(0)} - z \right) \left[ K_{x}^{(0)} K_{y}^{(0)} \right. \nonumber \\ &+& \left. \left( K_{x}^{(0)} - K_{y}^{(0)} \right) z \right]^{2} \,. \label{eq:supmat_eq_PG_clean_Bbar}
\end{eqnarray}
Notice that since the mapping \eref{supmat_eq_PG_define_z} is independent of $q$, both \eref{supmat_eq_PG_general_Abar} and \eref{supmat_eq_PG_general_Bbar} do not depend on the noise correlations in $q$.

\begin{figure}[t]
\setlength{\abovecaptionskip}{-2pt plus 3pt minus 2pt}
\begin{center}
\includegraphics[width=0.45\textwidth]{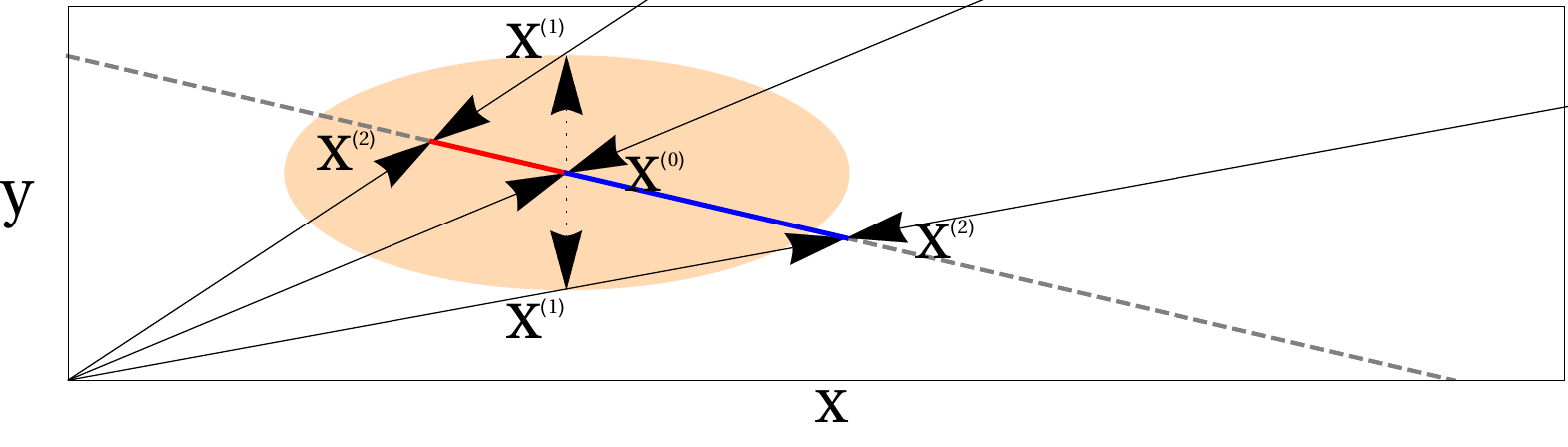}
\end{center}
\caption{Figure illustrating the origin of stochastically induced drift along the center manifold (CM). The gray dashed line shows the form of the deterministic center manifold, which intersects the $x$ axis at a higher value than the $y$ axis (phenotype $X$ has a higher carrying capacity due to the production of the public good). The red shaded circle illustrates the form of the Gaussian noise centered on the point $\bm{x}^{(0)}$ on the CM. Fluctuations in the population are equally likely to increase or decrease the frequency of the $Y$ phenotype to the points $\bm{x}^{(1)}$. Away from the CM, the deterministic pressure to the CM becomes prominent, forcing the system along quasi-deterministic trajectories back to the CM, at the points $\bm{x}^{(2)}$. The resulting distribution of $\bm{x}^{(2)}$ does not have a mean centered on $\bm{x}^{(0)}$. Rather, the distribution is shifted, inducing a drift in favor of the producing $X$ phenotype. }\label{fig:supmat_fig_projection_diagram}
\end{figure}

While the deterministic system features no dynamics along the CM, the effective SDE~(\ref{supmat_eq_quasineutral_effective_SDE}) does feature a drift in the mean state, embodied by $\mathcal{S}( z )$. Understanding the origin of this induced drift term requires considering the following. We envisage fluctuations arising from a single point on the CM, $\bm{x}^{(0)}$, which take to the system to a point off the CM, $\bm{x}^{(1)}$ (see \fref{fig:supmat_fig_projection_diagram}). The point $\bm{x}^{(1)}$ is clearly stochastic, but its distribution is approximately Gaussian, with a variance defined by $B(\bm{x}^{(0)})$. The fluctuation is now mapped back along a deterministic trajectory to a point $\bm{x}^{(2)}$ on the CM. The location $\bm{x}^{(2)}$ is also stochastic (dependent as it is on $\bm{x}^{(1)}$), and has its own distribution. The presence of the term $\mathcal{S}(z)$ in \eref{supmat_eq_quasineutral_effective_SDE} is indicative of the fact that the mean of the distribution of $\bm{x}^{(2)}$ is not $\bm{x}^{(0)}$; fluctuation events on average are mapped back to the CM with a preferred direction, inducing drift along the CM. Note that $\mathcal{S}(z)$ is positive along the length of the CM, which is defined on the interval $[0,K^{(0)}_{x}]$. 

We now turn our attention to the case when $\varepsilon > 0 $. So long as $\varepsilon$ is small, a separation of timescales is still present, though now no center manifold exists. Instead there is a slow manifold (SM), to which the deterministic system quickly relaxes, before slowly moving along it until phenotype $Y$ fixates. The equations for the population size at the boundaries of the SM are formally given by
\begin{eqnarray}
 N_{x} = R^{2} K_{x} \,, \quad  K_{x} = \left( \frac{  b_{x} \delta  }{ \delta \kappa - r p_{x} } \right) \equiv K^{(0)}_{x} + \mathcal{O}(\varepsilon) \,, \nonumber \\
 N_{y} = R^{2} K_{y} \,, \quad  K_{y} = \left( \frac{  b_{y} \delta  }{ \delta \kappa - r p_{y} } \right) \equiv K^{(0)}_{x} + \mathcal{O}(\varepsilon) \,. \nonumber \\ \label{eq:supmat_eq_NX_NY}
\end{eqnarray} 
In order to proceed with the stochastic calculation, we assume $\varepsilon \approx \mathcal{O}(R^{-2})$, and work order by order in $R^{-1}$. At leading order, the equation for the SM is identical to that of the CM, Eqs.~(\ref{supmat_eq_q_cm}) and (\ref{supmat_eq_y_cm}). The mapping to the SM is also unchanged at leading order from the quasi-neutral case (see \eref{supmat_eq_PG_define_z}). We proceed as before to obtain an effective description of the system dynamics in terms of $z$~\cite{parsons_rogers_2015}, now obtaining the dynamics,
\begin{equation}
 \dot{z} =  - \varepsilon \mathcal{D}( z ) + \frac{1}{R^{2}}\mathcal{S}( z ) + \frac{1}{R} \zeta (t) \,. \label{supmat_eq_nonneutral_effective_SDE}
\end{equation}
where
\begin{eqnarray}
 \mathcal{D}( z ) &=& - \left(   \frac{ d z }{ d x } A_{x}(\bm{x})  +   \frac{ d z }{ d y } A_{y}(\bm{x}) \right)|_{x=z,y=Z_{y}(z),q=Z_{q}(z)}  \,, \nonumber \\
                  &=& b z \left[ 1 - \left(\frac{\delta \kappa - p_{x} r }{ b \delta }\right) z \right]\,, \nonumber  \\
		  &=& \frac{b}{K_{x}^{(0)}} z \left( K_{x}^{(0)} - z \right) \,, \label{supmat_eq_PG_clean_AbarDet}
\end{eqnarray}
and $\mathcal{S}(t)$ and $\zeta(t)$ retain their form from the quasi-neutral case, Eqs.~(\ref{supmat_eq_PG_clean_Abar}) and (\ref{eq:supmat_eq_PG_clean_Bbar}). The function $\mathcal{D}(z)$ is the deterministic contribution to the dynamics along the SM. This expression is that which would be obtained using standard fast variable elimination techniques on the deterministic system. From \eref{supmat_eq_PG_clean_AbarDet}, we can see that $\mathcal{D}(z)$ is positive along the length of the SM and therefore acts (as we would expect) to increase the selective advantage of the non-producers, phenotype $Y$. There is therefore a conflict between the two components of the drift in the system. The term $\mathcal{D}(z)$ works against producers along the length of the SM, while $\mathcal{S}(z)$  creates a selective pressure in favor of producers. Ultimately, which term is more prevalent is dependent on the parameters $\varepsilon$ and $R$ (see \eref{supmat_eq_nonneutral_effective_SDE}); small $R$ leads to a small population size in which stochastic effects are stronger, and so producers are more likely to be selected for. In contrast, when the deterministic cost for good production is increased, the non-producers have an increased advantage over producers.

Adopting the notation used in the main text, in which we set $z=x$ (which is valid on the CM and SM at leading order), the expression for the SDE~(\ref{supmat_eq_nonneutral_effective_SDE}) can alternatively be written 
\begin{equation}
  \dot{x} =  \frac{b}{K_{x}^{(0)}} x \left( K_{x}^{(0)} - x \right) \left( \frac{ 1 }{ R^{2} } \,\mathcal{F}(x) - \varepsilon   \right)  + \frac{1}{R} \zeta(t) \,, \label{eq:supmat_eq_PG_maintext_effective}
\end{equation}
where 
\begin{equation}
 \mathcal{F}(x) = 2 \left( \frac{ K_{x}^{(0)} - K_{y}^{(0)} }{ (K_{x}^{(0)})^{2} (K_{y}^{(0)})^{2} } \right) \left[ K_{x}^{(0)} K_{y}^{(0)} + \left( K_{x}^{(0)} - K_{y}^{(0)} \right) x \right] \,.\label{eq:supmat_eq_Fx}
\end{equation}

\section{ Probability of fixation for the reduced public good model }\label{sec:supmat_sec_fixprob}

The fixation probability for a phenotype in a single variable system can be calculated using standard methods~\cite{gardiner_2009}. In order to conduct the calculation, we need expressions for the absorbing boundaries of the problem. For the reduced system given in \eref{supmat_eq_nonneutral_effective_SDE}, these lie at $z=0$ and $z=K^{(0)}_{x}$. The fact that the boundary for the problem exists at $z=K^{(0)}_{x}$, rather than $z=K_{x}$, is a consequence of the order to which we are working in $\varepsilon$. At this order the SM is approximated by the expression for the CM, which intersects the absorbing boundaries $x=0$ and $y=0$ at $z=0$ and $z=K^{(0)}_{x}$ respectively. Denoting $Q(z_{0})$ the fixation probability of producing phenotype $X$ given an initial frequency $z_{0}$ on the CM/SM, the fixation probability can be conveniently be expressed
\begin{eqnarray}
 Q(z_{0}) &=& \frac{ \int_{z=0}^{z_{0}} \psi(z) dz }{ \int_{ z=0 }^{ K_{x} } \psi(z) dz  } \,, \nonumber \\ 
\psi(z) &=& \exp \left[ \int_{0}^{z} \frac{2 ( - \varepsilon R \mathcal{D}( z' ) + \mathcal{S}( z' )) }{ \mathcal{B}(z') } dz'  \right] \,.
\end{eqnarray}
Substituting for $\mathcal{D}(z)$, $\mathcal{S}(z)$ and $\mathcal{B}(z)$ from Eqs.~(\ref{supmat_eq_PG_clean_AbarDet}),~(\ref{supmat_eq_PG_clean_Abar}) and (\ref{eq:supmat_eq_PG_clean_Bbar}), we find
\begin{eqnarray}
 Q(z_{0}) &=& \frac{1 - G(z_{0})}{ 1 - G(K_{x})  } \, ,  \nonumber \\ 
 G(z_{0}) &=&  \exp \left[ \frac{ \left( \varepsilon N_{y}^{(0)} K_{x}^{(0)} z_{0}  \right) }{ \left( K_{x}^{(0)} K_{y}^{(0)} + (K_{x}^{(0)} - K_{y}^{(0)}) z \right) } \right] \,. \label{supmat_eq_PG_fixprob_z}
\end{eqnarray}

The nature of these expressions can be understood more intuitively if we move from considering the initial frequency of $X$ on the CM, $z_{0}=n_{x0}/R^{2}$, to considering the initial fraction of phenotype $X$ on the CM, $f_{z0}$. The fraction and number of phenotype $X$ on the CM are related by
\begin{eqnarray}
 f_{z0} &=& \frac{ z }{ z + Z_{y}(z) } \,, \nonumber \\
 z &=&  \frac{ K_{x}^{(0)} K_{y}^{(0)} f_{z0} }{ K_{x}^{(0)} - ( K_{x}^{(0)} - K_{y}^{(0)} ) f_{z0} } \,.  
\end{eqnarray}
Substituting this into \eref{supmat_eq_PG_fixprob_z}, we find 
\begin{equation}
 Q(f_{z0}) = \frac{1 - \exp \left[ \varepsilon N_{y}^{(0)} f_{z0} \right] }{ 1 - \exp \left[ \varepsilon N_{y}^{(0)} \right]  } \, , \qquad Q(f_{z0})|_{\varepsilon = 0} = f_{z0} \,. \label{supmat_eq_PG_fixprob_f}
\end{equation}
On first appraisal, the fixation probabilities \eref{supmat_eq_PG_fixprob_f} appear to share the form of the well-mixed Moran model with weak selection. There is however one crucial distinction; the relation between $f_{z0}$ and $(x_{0},y_{0},q_{0})$ is dependent on the form of the CM/SM, and is not necessarily symmetric under the interchange of $X$ and $Y$. For instance, let us consider the quasi-neutral case ($\varepsilon=0$) with the population initially consisting of a mutant $X$ in a population of the $Y$ phenotype in its stationary state. Then $f_{z0} = 1/N_{y}$. In contrast, if the mutant is of phenotype $Y$, and the resident population consists of phenotype $X$ in the stationary state, $f_{z0} = 1 -  1/N_{x}$. Since $N_{x}$ and $N_{y}$ are distinct, these frequencies are not the same, and \eref{supmat_eq_PG_fixprob_f} is not symmetric under the interchange of phenotypes, undermining its apparent similarities with the Moran model.

In this section a crucial aspect of the selection reversal has been elucidated. The selection reversal along the SM is a result of the differing densities at which the populations of $X$ and $Y$ phenotypes reside in isolation. In a deterministic system, we would define the fitter phenotype as the one which fixates at long times. In stochastic Moran-type model, the fitter phenotype is defined as that with the greater invasion probability. Since Moran-type models feature a constant population size, $N$, the invasion probability of a mutant phenotype is defined by a unique initial condition; a single mutant, and $N-1$ residents. In systems such as the public good model discussed in this paper, the invasion probability is no longer defined uniquely by the specification of a single invading mutant; we must also define the size of the resident phenotype population and the public good density. If the system has been allowed to relax to a stationary state before the mutant is introduced, then selection reversal along the CM may be present, and it is possible for the producing phenotype to have a larger fixation probability than the non-producing phenotype. Thus the producing phenotype may be fitter.

\section{ Pairwise invasibility for non-producers, producers and hyper-producers }\label{sec:supMat_sec_pairwise}

In this section we explore the pairwise invasibility of three separate phenotypes, non-producers, producers and hyper-producers. We begin by noting that, under the assumption that the birth rates differ by only a small amount from phenotype to phenotype, the invasion probability of phenotype $i$ in a resident population $j$, $\phi_{i|j}$, can be expressed
\begin{eqnarray}
 \phi_{i|j} = \frac{1 - \exp\left[ (b_{i} - b_{j} ) / ( \kappa N_{j}^{(0)} )  \right] }{ 1 - \exp\left[ (b_{i} - b_{j} ) R^{2} / \kappa   \right] } \,.
\end{eqnarray}
We therefore define phenotype $i$ as fitter than phenotype $j$ if $\phi_{i|j}>\phi_{j|i}$. Let us now explicitly express the birth rates of each of the phenotypes as
\begin{eqnarray}
 \mathrm{Non-producer}: \, b_{y} &=& b \,, \nonumber \\
 \mathrm{Producer}: \, b_{x} &=& b(1-\varepsilon) \,, \nonumber \\
 \mathrm{Hyper-producer}:  \, b_{u} &=& b( 1 - a_{b} \varepsilon) \,. \nonumber \\ \label{supmat_eq_birth_pairwise}
\end{eqnarray}

\begin{figure}[t]
\setlength{\abovecaptionskip}{-2pt plus 3pt minus 2pt}
\begin{center}
\includegraphics[width=0.45\textwidth]{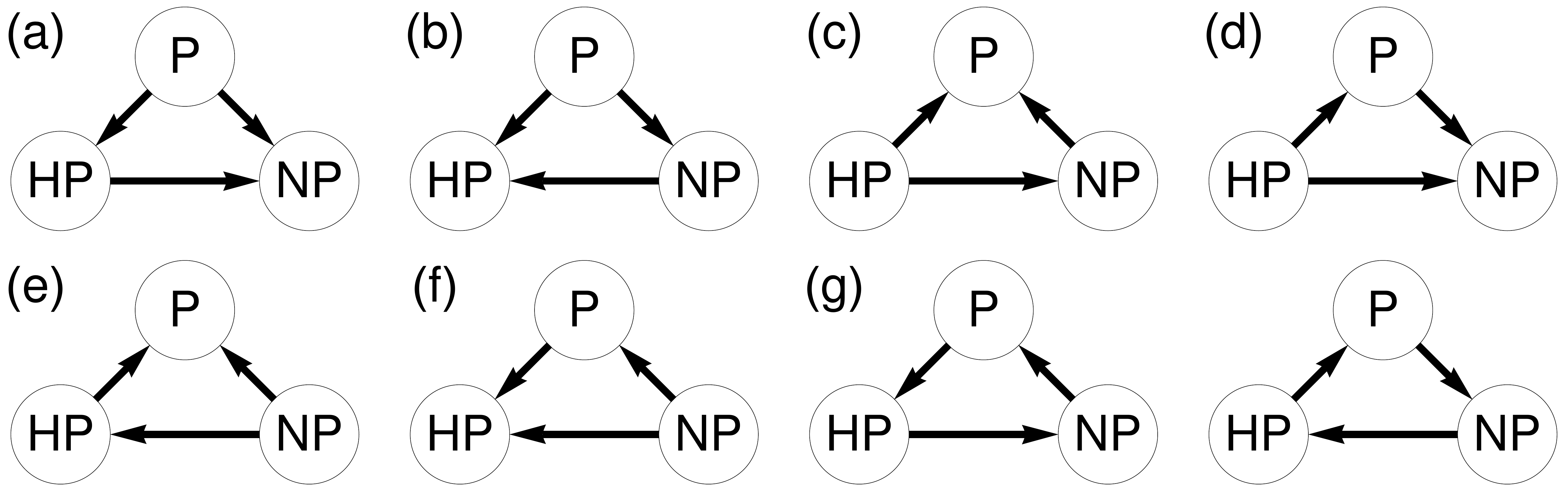}
\end{center}
\caption{ Eight different fitness rankings are possible based on the pairwise invasibility probabilities of non-producers, producers and hyper-producers.  (A) Producers have a larger invasion probability than both hyperproducers and nonproducers, while hyperproducers have a larger invasion probability than nonproducers. (B) Producers have a larger invasion probability than both hyperproducers and nonproducers, while nonproducers have a larger invasion probability than hyperproducers. (C) Hyperproducers have a larger invasion probability than both producers and nonproducers, while nonproducers have a larger invasion probability than producers. (D) Hyperproducers have a larger invasion probability than both producers and nonproducers, while producers have a larger invasion probability than producers. (E) Nonproducers have a larger invasion probability than both producers and hyperproducers, while hyperproducers have a larger invasion probability than producers. (F) Nonproducers have a larger invasion probability than both producers and hyperproducers, while producers have a larger invasion probability than hyperproducers. (G) Producers have a larger invasion probability than hyperproducers. Hyperproducers have a larger invasion probability than nonproducers. Nonproducers have a larger invasion probability than producers. (H) Producers have a larger invasion probability than nonproducers. Nonproducers have a larger invasion probability than hyperproducers. Hyperproducers have a larger invasion probability than producers. The nontransitive dynamics of G and H are not seen in the public good model. }\label{supmat_fig_invasion_scenarios}
\end{figure}

We now wish to obtain an expression for the critical costs to birth rate $\varepsilon$ at which producers are fitter than non-producers, hyper-producers are fitter than non-producers and hyper-producers are fitter than producers. To do this we must solve $\phi_{i|j}=\phi_{j|i}$ for $\varepsilon$ for each pair of phenotypes. An analytic solution is available if we set $\varepsilon = \tilde{\varepsilon}R^{-2}$ with $\tilde{\varepsilon}$ of order one, and expand Taylor expand in $R^{-2}$. Truncating at first order, we find that the critical cost for species $i$ to be fitter than species $j$, $\varepsilon_{i|j}$ is given by
\begin{eqnarray}
 \varepsilon_{i|j} = \frac{\kappa  \log \left[  ( p_{i} r - \delta  \kappa ) / ( p_{j} r - \delta  \kappa )\right] }{ \left[ (b_{i} - b_{j})/\varepsilon \right]  R^{2} } \,.
\end{eqnarray}
We note that this provides eight different possible scenarios of fitness ranking, described in \fref{supmat_fig_invasion_scenarios}. Substituting in our equations for the birth rates, \eref{supmat_eq_birth_pairwise}, these expressions become
\begin{eqnarray}
 \varepsilon_{x|y} &=& \frac{ \kappa }{ b R^{2} } \log \left[  -\frac{  \delta  \kappa }{ p_{x} r - \delta  \kappa } \right] \,, \\
 \varepsilon_{u|y} &=& \frac{ \kappa }{ a b R^{2} } \log \left[ - \frac{   \delta  \kappa }{ p_{u} r - \delta  \kappa }\right] \,,\\
 \varepsilon_{u|x} &=& \frac{ \kappa }{ (a-1) b R^{2} } \log \left[ \frac{  p_{x} r - \delta  \kappa }{ p_{u} r - \delta  \kappa }\right] \,.
\end{eqnarray}

Clearly the exact scenarios which emerge for a given set of parameters depends on the relationship between $p_{x}$ and $p_{u}$. We make the assumption 
\begin{eqnarray}
\quad p_{u} = a_{p} p_{x} \,.
\end{eqnarray}
For $a_{p}>a_{b}$, the hyper-producer pays a discounted cost to its birth rate for its additional good production. In this situation, only scenarios (c-f) are possible in \fref{supmat_fig_invasion_scenarios}.  It is always better to be a hyper-producer or a non-producer, depending on the production rate $p_{x}$ and $\epsilon$. This `all or nothing' result makes intuitive sense; if the hyper-producer produces much more than the producer, but pays only fractionally more to its birth rate, any region in which production is favored will be disproportionately advantageous to the hyper-producers. In contrast, if $a_{p}<a_{b}$, the hyper-producer receives decreasing production returns as a function of the cost it pays to birth in comparison with the producer. In this case, scenarios (a-b) and (e-f) are possible. Either producers or non-producers are favored, and hyper-producers are never favored.

\section{Generality of results}\label{sec:supmat_sec_general_results}

We begin by specifying in a very general way the dynamics of an arbitrary IBM with $m$ distinct types of constituent, fully described by a set of $u$ reaction rates. The model can be expressed in chemical reaction notation as
\begin{eqnarray}\label{supmat_generalReaction}
 \sum_{i=1}^{m} a_{\mu i}X_{i} \xrightarrow{ r_{\mu} } \sum_{i=1}^{m}b_{ \mu i}X_{i}, \quad \forall \mu=1, \dots u,
\end{eqnarray}
where $a_{\mu i}$ and $b_{ \mu i}$ respectively specify the reactants and products of the $\mu^{th}$ reaction, and $r_{\mu}$ are the reaction rate constants (see, for example, \eref{supmat_eq_reactions}). The stoichiometric matrix is defined by $\nu_{i \mu } = b_{\mu i}-a_{\mu i}$, whose elements give the change in number of the $i^{\mathrm{th}}$ species due to the $\mu^{\mathrm{th}}$ reaction. Together with the rate constants $r_{\mu}$, the stoichiometric matrix allows us to express the transition rates
\begin{eqnarray}
 T_{\mu}( \bm{n} + \bm{\nu}_{\mu} |\bm{n}) = r_{\mu}  \prod_{i=1}^{m} a_{\mu i} \frac{n_{i}}{R^{2}} \,, \label{supmat_relate_reactants_transitions}
\end{eqnarray}
where $R^{2}$ once again is a controlled measure of how often constituents interact (see \eref{supmat_eq_trans}). In the well-mixed model, it therefore directly controls the typical area of the system. Together with the master equation~(\ref{eq:supmat_meqn_general}), the full stochastic dynamics are specified.

With a general notation now in hand, we now begin to define the specific type of system we will analyze. We consider a system consisting of two phenotypes, $X_{1}$ and $X_{2}$, who interact with a set of discrete ecosystem variables $X_{i}$, for $i = 3, \ldots , N$. The state of the system at any time is given by the number of each phenotype and ecosystem constituent $\bm{n}=(n_{1},n_{2},n_{3},\ldots,n_{N})$. The situation we envisage is as follows; while the interplay between the phenotypes and the ecosystem is relevant for the dynamics, we are primarily interested in the evolutionary dynamics and outcome of competition between the two phenotypes. We make the following assumptions on their dynamics;
\begin{enumerate}
 \item Each phenotype birth and death event is proportional to the number of that phenotype;
\begin{eqnarray}
\mathrm{if} \quad \nu_{1\mu} &\neq& 0 \quad \mathrm{then} \quad a_{\mu1}>0 \,,  \qquad  \mathrm{and} \nonumber \\  \mathrm{if} \quad \nu_{2\mu} &\neq& 0  \quad \mathrm{then} \quad a_{\mu2}>0 \,.
\end{eqnarray}

 \item The phenotypes are very similar in their utilization of the ecosystem. For each $\mu^{th}$ reaction that changes the frequency of $X_{1}$, there therefore exists a similar reaction $\mu'$ that changes the frequency of $X_{2}$ such that;
\begin{equation}
 \nu_{1 \mu} r_{\mu} = \nu_{2 \mu' } ( r_{\mu'} + \mathcal{O}(\varepsilon) ) \,.
\end{equation}

 \item There is no reaction which simultaneously changes the frequencies of the phenotypes (i.e. no cannibalization or simultaneous killing);
\begin{equation}
 \nu_{1 \mu} \nu_{2 \mu } = 0 \qquad \forall \, \mu \,.
\end{equation}

\end{enumerate}
The phenotypes may however differ significantly in their effect on the ecosystem, so that one phenotype may deplete or increase ecosystem constituents in an entirely distinct way to the other (for instance, the production of a public good by phenotype $X$ in \eref{supmat_eq_reactions}).

As $R$ is increases so too does the number of each phenotype and ecosystem constituent. If $R$ is sufficiently large, once again a system-size expansion of the master equation can be conducted. Making the change of variables $x_{1} = n_{1}/R^{2}$, $x_{2} = n_{2} /R^{2} $ and $e_{i}=n_{i-2}/R^{2}$, we obtain the set of It\={o} SDEs
\begin{eqnarray}
 \frac{\mathrm{d}x_{1}}{\mathrm{d}t} &=& x_{1} \left[ F^{(0)}(\bm{x},\bm{e}) - \varepsilon F^{(1)}(\bm{x},\bm{e}) \right] + \frac{1}{R}\eta_{1}(t) \, , \nonumber \\
 \frac{\mathrm{d}x_{2}}{\mathrm{d}t} &=& x_{2}  F^{(0)}(\bm{x},\bm{e}) + \frac{1}{R}\eta_{2}(t)  \, , \nonumber \\
 \frac{\mathrm{d}e_{i}}{\mathrm{d}t} &=& h_{i}(\bm{x},\bm{e}) + \frac{1}{R}\beta_{i}(t) \,, \quad \forall \, i = 1 , \ldots N \,. \label{supmat_eq_general_SDEs}
\end{eqnarray}
The deterministic contribution to the SDEs can be determined from the transitions via
\begin{eqnarray}
 x_{1} \left[ F^{(0)}(\bm{x},\bm{e}) \right. &-& \left. \varepsilon F^{(1)}(\bm{x},\bm{e}) \right] = \nonumber \\ &&  \sum_{\mu=1}^{u} \nu_{1 \mu}T_{\mu}\left[ R^{2} (\bm{x},\bm{e})^{T} + \bm{\nu}_{\mu} | (\bm{x},\bm{e})^{T} \right] \,, \label{supmat_eq_general_drift_1} \\
 x_{2} F^{(0)}(\bm{x},\bm{e}) &=&  \sum_{\mu=1}^{u} \nu_{2 \mu}T_{\mu}\left[ R^{2} (\bm{x},\bm{e})^{T} + \bm{\nu}_{\mu} | (\bm{x},\bm{e})^{T} \right] \,,  \label{supmat_eq_general_drift_2} \\
 h_{i}(\bm{x},\bm{e}) &=& \sum_{\mu=1}^{u} \nu_{(i+2) \mu}T_{\mu}\left[ R^{2} (\bm{x},\bm{e})^{T} + \bm{\nu}_{\mu} | (\bm{x},\bm{e})^{T} \right] \,. \nonumber
\end{eqnarray}
Notice that the relationship between Eqs.~(\ref{supmat_eq_general_drift_1}) and (\ref{supmat_eq_general_drift_2}) is controlled by assumption 2. The correlations in the noise meanwhile are given by 
\begin{eqnarray}
&& \langle \eta_{1}(t) \eta_{1}(t') \rangle = \nonumber \\ && \delta(t-t') \lim_{\varepsilon \to 0 } \sum_{\mu=1}^{u} \nu_{1 \mu}^{2} T_{\mu}\left[ R (\bm{x},\bm{e}) + \bm{\nu}_{\mu} | (\bm{x},\bm{e}) \right] \,, \label{supmat_eq_general_phenotype_correlation_1}\\
&& \langle \eta_{2}(t) \eta_{2}(t') \rangle = \nonumber \\ && \delta(t-t') \lim_{\varepsilon \to 0 } \sum_{\mu=1}^{u} \nu_{2 \mu}^{2} T_{\mu}\left[ R (\bm{x},\bm{e}) + \bm{\nu}_{\mu} | (\bm{x},\bm{e}) \right] \,, \label{supmat_eq_general_phenotype_correlation_2}\\
&& \langle \eta_{1}(t) \eta_{2}(t') \rangle =  0 \,, \label{supmat_eq_general_phenotype_correlation_3}\\
&& \langle \eta_{i}(t) \beta_{j}(t') \rangle = \nonumber \\ && \delta(t-t')\lim_{\varepsilon \to 0 } \sum_{\mu=1}^{u} \nu_{i \mu}\nu_{(j+2) \mu} T_{\mu}\left[ R (\bm{x},\bm{e}) + \bm{\nu}_{\mu} | (\bm{x},\bm{e}) \right] \,, \\
 &&\langle \beta_{i}(t) \beta_{j}(t') \rangle = \nonumber \\ && \delta(t-t') \lim_{\varepsilon \to 0 } \sum_{\mu=1}^{u} \nu_{(i+2) \mu}\nu_{(j+2) \mu} T_{\mu}\left[ R (\bm{x},\bm{e}) + \bm{\nu}_{\mu} | (\bm{x},\bm{e}) \right] \,, \nonumber \\
\end{eqnarray}
at leading order in $\varepsilon$. The lack of noise correlation between the phenotypes, \eref{supmat_eq_general_phenotype_correlation_3}, is a consequence of assumption 3. Assumption 2 allows us to rewrite Eqs.~(\ref{supmat_eq_general_phenotype_correlation_1}) and (\ref{supmat_eq_general_phenotype_correlation_2}) as
\begin{eqnarray}
  \langle \eta_{1}(t) \eta_{1}(t') \rangle &=& \delta(t-t') x_{1} H(\bm{x},\bm{e}) \,, \nonumber\\
  \langle \eta_{2}(t) \eta_{2}(t') \rangle &=& \delta(t-t') x_{2} H(\bm{x},\bm{e}) \,. \label{supmat_eq_phenotype_correlation_rule}
\end{eqnarray}
An example of a system where this condition is not enforced is explored in \sref{sec:supmat_sec_comp_models}.

To begin our analysis of the SDEs, a quasi-neutral limit is considered in which $\varepsilon=0$. Then the deterministic ODEs for the system (the SDEs in the limit $R\rightarrow \infty$) lead to a manifold of fixed points associated with the focus phenotypes. We now make two additional assumptions; 
\begin{enumerate}
\setcounter{enumi}{4}
 \item There exits a single stable, well behaved, manifold
 \item This manifold is one-dimensional, and so can be paramaterized by a single variable
\end{enumerate}
We then choose to parameterize the manifold in terms of $x_{1}$, which for clarity we label $z$ on the CM. The CM is then defined by the set of equations
\begin{equation}
x_{1} = z \,, \qquad x_{2} = Z_{2}(z) \, ,\qquad e_{i} = Z_{ei}(z) \quad \forall \, i=2, \ldots N  \,. \label{supmat_eq_general_CM}
\end{equation}
The system dynamics are now entirely analogous to that of the public good model in the quasi-neutral limit. Deterministically, the system comes to rest at a point on the CM (which depends on the system's initial conditions) at which it stays indefinitely, and when stochasticity is included the system moves along the CM until one of the phenotypes fixates. A timescale separation is present so long as the composition of the population changes on a slower timescale to that of the collapse to the CM. In practice, the timescale of the collapse to the CM can be inferred from the eigenvalues of \eref{supmat_eq_general_SDEs} linearised about the CM. The magnitude of the smallest non-zero eigenvalue is indicative of the slowest component of collapse to the CM~\cite{constable_2013}. This should be much larger than the timescale at which the system moves along the CM, which is of order $R^{-1}$~\cite{constable_2014_phys}.

In order to implement the timescale separation, a non-linear projection is applied to the system which maps fluctuations back to the CM. This can be seen to be equivalent to transforming into the deterministically invariant variable whose existence is guaranteed by the existence of the CM~\cite{arnold_2003}, setting the dynamics in all other variables equal to zero, and evaluating the variables themselves on the CM. What form does this mapping take, in the quasi-neutral limit, for \eref{supmat_eq_general_SDEs}? Since the dynamical equations for the phenotypes take on the form of degenerate replicator equations in the limit $\varepsilon \rightarrow 0$, the ratio $x_{1}/x_{2}$ is deterministically invariant, regardless of the other parameters. Therefore the non-linear mapping may be obtained by solving the following equation for $z$;
\begin{equation}
 \frac{ z }{ Z_{2}( z ) } = \frac{ x_{1} }{ x_{2} } \, , \quad \rightarrow \quad  z = Y(x_{1},x_{2}) \,. \label{supmat_eq_general_projection}
\end{equation}
The resulting effective description for the quasi-neutral system on the CM can be denoted
\begin{equation}
\dot{z} =  \frac{1}{R}\mathcal{S}( z ) + \frac{1}{\sqrt{R}} \zeta(t) \,.
\end{equation}
Note that while the deterministic system evaluated on the CM had no drift dynamics, the reduced system may. Mathematically, this is a consequence of the fact that the equations are defined strictly in the It\={o} sense (from the underlying IBM) and therefore the normal rules of calculus do not apply. Instead, any nonlinear transformation induces a drift, in general given by
\begin{eqnarray}
 \mathcal{S}( z ) &=& \frac{1}{2} \left. \left[ \sum_{ij}^{2} \left( \frac{\partial z }{\partial x_{i} \partial x_{j} } B_{ij} \right) \right. \right. \nonumber \\ &+& \left. \left. \sum_{ij}^{N} \left( \frac{\partial z }{\partial e_{i} \partial e_{j} } B_{eij} \right)  \right] \right|_{x_{1}=z,x_{2}=Z_{2}(z),e_{i}=Z_{ei}(z)} . \label{supmat_eq_Abar_stoch_1}
\end{eqnarray}
However, since the mapping $z$ is independent of the ecosystem variables $\bm{e}$ (see \eref{supmat_eq_general_projection}), \eref{supmat_eq_Abar_stoch_1} can be simplified to 
\begin{equation}
 \mathcal{S}( z ) = \frac{1}{2} \left.\sum_{ij}^{2} \left( \frac{\partial z }{\partial x_{i} \partial x_{j} } B_{ij}\right) \right|_{x_{1}=z,x_{2}=Z_{2}(z),e_{i}=Z_{ei}(z)} \,. \label{eq:supmat_eq_Abar_stoch_2}
\end{equation}
The form of the correlations in $\zeta(t)$ are now given by
\begin{equation}
\mathcal{B}(z) = \left. \sum_{ij}^{2} \left(\left[ \frac{\partial z }{\partial x_{i} } \right]_{i} \left[ \frac{\partial z }{ \partial x_{j} }\right]_{j} B_{ij}(\bm{x}) \right) \right|_{x=z,x_{2}=Z_{2}(z),e_{i}=Z_{ei}(z)} \,, \label{eq:supmat_eq_general_Bbar}
\end{equation}
where once again we have taken advantage of the property $(dz/de_{i})=0$ for all $i$.

In this very general scenario, what inferences can we make about $\mathcal{S}( z )$? To answer this, it is convenient to return to our original SDEs, \eref{supmat_eq_general_SDEs}, and implement the timescale separation in a different fashion. We begin by transforming into variables measuring the total size of the $x_{1}$ and $x_{2}$ population and the fraction of type $x_{1}$;
\begin{eqnarray}
 N_{T} = x_{1} + x_{2} \,, \quad f_{1} = \frac{ x_{1} }{ x_{1} + x_{2} } \,, \nonumber \\  \rightarrow \quad x_{1} = f_{1} N_{T} \,, \quad x_{2} = N_{T}(1 - f_{1}) \,.
\end{eqnarray}
Applying this transformation, taking care to account for the impact of It\={o} calculus, we arrive at the following SDEs for the system;
\begin{eqnarray}
 \frac{\mathrm{d}f_{1}}{\mathrm{d}t} &=& \frac{1}{2R^{2}}\sum_{i,j=1}^{2}\frac{\partial^{2} f_{1} }{\partial x_{i} \partial x_{j} } B_{ij} + \frac{1}{R}\tilde{\eta}_{1}(t) \, , \nonumber \\
 \frac{\mathrm{d}N_{T}}{\mathrm{d}t} &=& N_{T} F^{(0)}(\bm{x},\bm{e})   +     \frac{1}{R}\tilde{\eta}_{2}(t) \, , \nonumber \\
 \frac{\mathrm{d}e_{i}}{\mathrm{d}t} &=& h_{i}(\bm{x},\bm{e}) + \frac{1}{R} \tilde{\beta_{i}}(t) \,, \quad \forall \, i = 1 , \ldots N \,.
\end{eqnarray}
By conducting the transformation, we immediately notice a few things. Most trivially, the forms of the noise correlations are now altered in all variables. Second, since the transformation into the variable $N_{T}$ was linear, its governing SDE contains no noise-induced elements. Finally, the non-linear transformation into $f_{1}$ has resulted in a noise induced drift term. This drift term however is only dependent on the noise correlation structure between $x_{1}$ and $x_{2}$. Evaluating the dynamics for $N_{T}$ and $\bm{e}$ on the CM and substituting in the remaining expressions from Eqs.~(\ref{supmat_eq_general_phenotype_correlation_3}) and (\ref{supmat_eq_phenotype_correlation_rule}), we obtain the following one-dimensional SDE for $f_{1}$;
\begin{equation}
 \frac{ \mathrm{d} f_{1} }{ \mathrm{d} t } = \frac{1}{R}\tilde{\eta}_{1}(t) \,,  \label{eq_f1_neutral}
\end{equation}
where $\tilde{\eta}_{1}(t)$ is evaluated on the CM. There are no deterministic dynamics in our reduced dimension description of $f_{1}$. This is a consequence of assumptions 2 and 3. The equation for the fixation probability of phenotype $X_{1}$ given an initial \emph{fraction} $f_{10}$ on the CM, $Q(f_{10})$, is then, regardless of the noise form,
\begin{eqnarray}
 Q(f_{10}) = f_{10} \,.
\end{eqnarray}
Crucially however, $f_{1}$ is evaluated on the CM, which may vary depending on the constitution of the population;
\begin{eqnarray}
 f_{10} = \frac{ x_{10} }{ x_{10} + Z_{2}(x_{10})  } \,.
\end{eqnarray}
If $\left[ d Z_{2}(x_{10}) /dx_{10} \right] < 1 $, then the total phenotype population decreases with increasing $x_{20}$, and phenotype $X_{1}$ has a larger invasion probability than $X_{2}$. From this we can infer that $\mathcal{S}(z)$ will be positive on average along the length of the CM;
\begin{eqnarray}
 \int_{z=0}^{N_{1}/R^{2}} \mathcal{S}(z) dz > 0 \,. \label{eq:supmat_integral_S}
\end{eqnarray}
Therefore, the phenotype with the higher carrying capacity will be stochastically selected for in this quasi-neutral case, regardless of their interaction with the environment. We note once again that this result is in general dependent on assumption 2. If assumption 2 does not hold then there will be correlations between the fluctuations $\eta_{1}(t)$ and $\eta_{2}(t)$ and, rather than the equation for the time evolution of $f_{1}$ featuring no mean drift (as in Eq.~(\ref{eq_f1_neutral}) there will be a noise induced drift term favoring one or other of the phenotypes. The exact form of this term will be highly dependent on the exact form of the interactions between the phenotypes, a full treatment of which lies outside the scope of this paper.

Now suppose that $\varepsilon>0$, so that the system is non-neutral. Now there exists no CM. There is no line of deterministic fixed points, and therefore no invariant variable to project our variables on to and reduce the problem. However, under the assumption that $\varepsilon$ is small there is still a separation of timescales. If $\varepsilon$ is sufficiently small, the slow manifold (and the projection to it) can be approximated by the results from the quasi-neutral case (see Eqs.~(\ref{supmat_eq_general_CM}) and (\ref{supmat_eq_general_projection})), plus an $\varepsilon$ correction. A perturbative analysis can thus be conducted, and, under the assumption the $\varepsilon \approx \mathcal{O}(R^{-2})$, at leading order we have
\begin{eqnarray}
 \dot{z} =  - \varepsilon \mathcal{D}(z) + \frac{1}{R^{2}}\mathcal{S}( z )  + \frac{1}{R} \bar{\eta}(t) \,. \label{supmat_eq_Abar_stoch_det_SDE}
\end{eqnarray}
The form of $\mathcal{S}( z )$ is unchanged from \eref{supmat_eq_Abar_stoch_1}, while the new deterministic contribution to the drift takes the form
\begin{equation}
 \mathcal{D}(z) = - \left. \sum_{i=1}^{N}\left( \frac{ d z }{ d x_{i} } \frac{ d x_{i} }{ d t } \right) \right|_{x_{1}=z,x_{2}=Z_{2}(z),\bm{e}=\bm{Z_{e}}(z) } \,.
\end{equation}
Once again however, the projection is simply a function of $x_{1}$ and $x_{2}$, and so
\begin{eqnarray}
 \mathcal{D}(z) &=& - \left. \left( x_{1} F^{(0)}(\bm{x}) \frac{ d z }{ d x_{1} } +  x_{2} F^{(0)}(\bm{x}) \frac{ d z }{ d x_{2} }  
\right. \right. \nonumber \\ &-& \left. \left. \varepsilon x_{1} F^{(1)} \frac{ d z }{ d x_{1} } \right) \right|_{x_{1}=z,x_{2}=Z_{2}(z),\bm{e}=\bm{Z_{e}}(z) } \,.
\end{eqnarray}
Finally, we also know that in the limit $ \varepsilon \rightarrow 0$ this deterministic contribution to the dynamics on the CM, $\mathcal{D}(z)$, should disappear. Therefore the first two terms in the  above equation must cancel, leaving us with 
\begin{equation}
 \mathcal{D}(z) = \left. \varepsilon z \left(F^{(1)}(\bm{x}) \frac{ d z }{ d x_{1} } \right) \right|_{x_{1}=z,x_{i}=Z_{i}(z)} \,. \label{eq:supmat_eq_general_Abar_det}
\end{equation}
We now have a much simpler system to deal with. Say that $F^{(1)}( \bm{x} )$ is strictly positive. Then this will be a term which consistently decreases the value of $x_{1}$. Based on physical arguments, we would expect that, regardless of the form of $\zeta$, $\mathcal{D}(z)$ must be positive. We still require the exact form of $z$ (see \eref{supmat_eq_general_projection}) to make analytic progress and specific predictions. Generally however, we have shown that $\mathcal{S}( z )$ will be positive so long as species $X_{1}$ has a larger carrying capacity (subject to the above conditions). A consideration of \eref{supmat_eq_Abar_stoch_det_SDE} shows that even when the system is non-neutral, for sufficiently weak selection/small $R$, there will be a tradeoff between stochastic `strength in numbers' and deterministic costs for high-density behavior.

\section{Illustrating generality with reference to a complimentary systems: The stochastic Lotka-Volterra system }\label{sec:supmat_sec_comp_models}

In \sref{supmat_sec_public_good_reduction} it was noted that deterministically the public good model reduces to a competitive Lotka-Volterra model under the elimination of the fast public good dynamics. However, it is important to note that though they may be deterministically equivalent at long times, due to alterations in the demographic noise structure the two systems have distinct behaviors. Despite this, the qualitative picture remains the same; for the quasi-neutral system, the fixation probability of each type is simply proportional to its initial fraction in the population, while when selection is introduced, there is playoff between stochastic and deterministic effects. To illustrate this, we investigate the stochastic Lotka-Volterra competition model (SLVC), derived from first principles.

In this section we analyze a stochastic Lotka-Volterra competition model using the methods developed in \sref{sec:supmat_sec_general_results}. We assume a population composed of two phenotypes, $X_{1}$ and $X_{2}$, whose numbers in the system are measured by $\bm{n}=(n_{1},n_{2})$. The phenotypes are born, die and compete with each other. In particular, we define the system to be governed by the probability transition rates
\begin{eqnarray}\label{supmat_eq_SLVC_trans_rates}
&T_1(n_1 + 1,n_2|n_1,n_2) = b_1 n_1 \, , &\nonumber \\
&T_2(n_1 - 1,n_2|n_1,n_2) = d_1 n_1 + \frac{ c_{1} }{ R^{2} } n_1^{2} + \frac{ c_{2} }{ R^{2} } n_{1} n_{2}  \, , & \nonumber \\
&T_3(n_1,n_2 + 1|n_1,n_2) = b_2 n_2\,, &\nonumber \\
&T_4(n_1,n_2 - 1|n_1,n_2) = d_2 n_2 + \frac{ c_{1} }{ R^{2} } n_1 n_2 + \frac{ c_{2} }{ R^{2} } n_{2}^{2} \, . &\nonumber
\end{eqnarray}
Together with \eref{eq:supmat_meqn_general}, this fully specifies the stochastic dynamics. Taking the limit of large $R$, we can once again obtain a mesoscopic description of the system;
\begin{eqnarray}
\frac{\mathrm{d}x_{1}}{\mathrm{d} t } &=& x_{1}\left( (b_{1} - d_{1} ) - c_{1} x_{1} - c_{2} x_{2} \right) + \frac{1}{ R } \eta_{1}( t ), \nonumber \\
\frac{\mathrm{d}x_{2}}{\mathrm{d} t } &=& x_{2}\left( (b_{2} - d_{2} ) - c_{1} x_{1} - c_{2} x_{2} \right) + \frac{1}{ R } \eta_{2}( t ), \nonumber \\
\label{supmat_eq_SLVC_SDE_orig}
\end{eqnarray}
where $\eta_{i}(t)$ have correlation structure \eref{supmat_eq_correlations} with $B_{ij}(\bm{x})$ term given by
\begin{eqnarray}
B_{11}(\bm{x}) &=& x_{1}\left( ( b_{1} + d_{1} ) + c_{1} x_{1} + c_{2} x_{2} \right) \,, \nonumber \\
B_{22}(\bm{x}) &=& x_{2}\left( ( b_{2} + d_{2} ) + c_{1} x_{1} + c_{2} x_{2} \right) \,, \nonumber \\
B_{12}(\bm{x}) &\equiv& B_{21}(\bm{x}) = 0 \,.
\end{eqnarray}
Note that the noise structure is \emph{not} the same as that in \eref{supmat_eq_phenotype_correlation_rule}; two phenotypes with an equal effective reproduction rate $b_{1}-d_{1}=b_{2}-d_{2}$ have the same deterministic fitness, but distinct multiplicative noise. Phenotypes which are reproducing and dying more quickly are subject to greater noise as they have a larger rate of population turnover. We will however proceed to consider this more general scenario in order to illustrate what can happen when this assumption is not enforced. Finally, we impose a separation of timescales by setting
\begin{eqnarray}
 b_{1} - d_{1} = \tilde{b}( 1 - \varepsilon ) \,, \quad b_{2} - d_{2} = \tilde{b} \,.
\end{eqnarray}
A CM thus exists if $\varepsilon = 0$, and an SM while $\varepsilon$ is small. The parameter $\tilde{b}$ is an effective birth rate encompassing birth and death, while $\varepsilon$ is a fitness cost paid by phenotype $X_{1}$ either in terms of a decreased birth rate, or increased death rate, relative to phenotype $X_{2}$.

In the case $\varepsilon = 0$, the system is quasi-neutral, and so a CM exists. The equation for the CM $x_{2}=Z_{2}(x_{1})$ (see \eref{supmat_eq_general_CM}) and its intersection with the boundaries $x_{2}=0$ and $x_{1}=0$, $K_{1}^{(0)}$ and $K_{2}^{(0)}$ respectively, are
\begin{eqnarray}
 Z_{2}(x_{1}) = \frac{1}{c_{2}} \left( \tilde{b} - c_{1} x_{1} \right) \,, \nonumber \\ \quad  K_{1}^{(0)} = \frac{ \tilde{b} }{ c_{1} } \,, \quad K_{2}^{(0)} = \frac{ \tilde{b} }{ c_{2} } \,.
\end{eqnarray}
The parameters $K_{1}^{(0)}$ and $K_{2}^{(0)}$ give the frequency of $X_{1}$ and $X_{2}$ phenotypes in isolation. We assume that $c_{2}>c_{1}$ and thus that phenotype $X_{1}$ exists at higher densities than phenotype $X_{2}$. Finally, the mapping from any point $(x_{1},x_{2})$ to a coordinate $z=x_{1}$ on the CM is determined from \eref{supmat_eq_general_projection};
\begin{eqnarray}
 z = \frac{ \tilde{b} x_{1} }{ c_{1} x_{1} + c_{2} x_{2} } \,.
\end{eqnarray}
We can now obtain expressions for $\mathcal{D}( z )$, $\mathcal{S}( z )$ and $\mathcal{B}(z)$ directly from Eqs.~(\ref{eq:supmat_eq_general_Abar_det}), (\ref{supmat_eq_Abar_stoch_1}) and (\ref{eq:supmat_eq_general_Bbar});
\begin{eqnarray}
 \mathcal{D}( z ) &=&  -   z \left( \tilde{b} - c_{1} z\right)  \,,\\
 \mathcal{S}( z ) &=& \frac{2}{ \tilde{b}^{2} }  z \left( \tilde{b} - c_{1} z\right) \left( c_{2} (\tilde{b} + d_{2} ) - c_{1} (\tilde{b} + d_{1} ) \right) \,, \\
 \mathcal{B}(z) &=&   \frac{2}{ \tilde{b}^{2} }  z \left( \tilde{b} - c_{1} z\right) \left[ z  \left( c_{2} (\tilde{b} + d_{2} ) \right. \right. \nonumber \\ &-& \left. \left. c_{1} (\tilde{b} + d_{1} ) \right) + \tilde{b} ( d_{1} + \beta)\right]  \,.
\end{eqnarray}
The equation can now be solved to calculate the fixation probability of phenotype $X_{1}$ along the CM. In terms of the initial fraction of $X_{1}$, $f_{1}$, we find
\begin{eqnarray}
 Q(f_{1}) &=& \frac{1 - \chi( f_{1} ) }{ 1 - \chi( 1 ) } \,, \nonumber \\
 \chi(f_{1}) &=& \left[ \left( \frac{ K_{1}^{(0)} }{ d_{1} + \tilde{b} } \right) \left( \frac{ d_{1}( 1 - f_{1} ) + d_{2} f_{1} + \tilde{b} }{ K_{1}^{(0)}(1-f_{1}) + f_{1} K_{2}^{(0)} } \right) \right] ^{ - \theta }\,,
\end{eqnarray}
where $\theta$ is a parameter given by
\begin{eqnarray}
 \theta = \left( 1 + \frac{ K_{1}^{(0)} K_{2}^{(0)} R^{2} \tilde{b} \varepsilon }{ K_{2}^{(0)} (d_{1} + \tilde{b} ) - K_{1}^{(0)} ( d_{2} + \tilde{b} ) } \right) \,.
\end{eqnarray}

Let us consider the special case $\varepsilon = 0$. The fixation probability then becomes
\begin{eqnarray}
 Q(f_{1})|_{\varepsilon=0} = \frac{  f_{1} ( d_{2} + \tilde{b} ) }{ d_{1}( 1 - f ) + d_{2} f + \tilde{b} } \,.
\end{eqnarray}
The species with the lower death rate (and death rate, since $\tilde{b}$ is fixed), has a greater probability of fixation than the species with the higher birth rate/death rate. This insight, made in  \cite{parsons_quince_2007_1,doering_2012}, is a result of the higher levels of noise experienced by the phenotype with the high birth and death rates. This makes it easier for the longer lived phenotype, (lower birth/death rates), to invade and fixate. For the purposes of this paper, we ignore such effects in order to focus on systems in which the carrying capacity of the phenotypes alone is responsible for the differences in noise experienced by the phenotypes on the CM/SM.

To this end, we now focus on the case $b_{1}=b_{2}\equiv b$, $d_{1}=d_{2}\equiv d$. In this case,  $Q(f_{1})|_{\varepsilon=0} = f_{1}$, and $Q(f_{1})$ in general becomes
\begin{eqnarray}
 Q(f_{1}) &=& \frac{1 - \chi( f_{1} ) }{ 1 - \chi( 1 ) } \,, \nonumber \\
\chi(f_{1}) &=&  \left( \frac{ K_{1}^{(0)} } { K_{1} (1 -f_{1} ) +f_{1}K_{2}^{(0)} } \right)^{-\theta}    \,,
\end{eqnarray}
where $\theta$ is now given by 
\begin{eqnarray}
 \theta = \left( 1 + \frac{ K_{1}^{(0)} K_{2}^{(0)} R^{2} ( b - d ) \varepsilon }{ ( K_{2}^{(0)} - K_{1}^{(0)} ) b  } \right) \,.
\end{eqnarray}

The invasion probabilities $\phi_{1}$ and $\phi_{2}$ meanwhile are given by
\begin{eqnarray}
 \phi_{1} = Q(N_{2}^{-1}) \,,\quad \phi_{2} = 1 - Q( 1 - N_{1}^{-1} ) \,.
\end{eqnarray}
We can use the above expressions to obtain an approximate value for the maximum cost to birth rate that can be paid in order that selection reversal is observed. Assuming $N_{1}^{-1}$ and $N_{2}^{-1}$ are of order $\varepsilon$ and Taylor expanding in $\varepsilon$, we find the cost to birth must obey 
\begin{equation}
 \frac{1}{N_{2}} \left( \frac{ b }{ b - d } \right) \left( 1 - \frac{ N_{2} }{ N_{1} } \right) > \varepsilon
\end{equation}
for the direction of selection to be reversed. This is analogous to Eq.~($8$) in the main text.

\section{Order of magnitude parameter estimates}\label{sec:supmat_sec_params}

In this section we seek an illustrative set of parameters for use in the model in order emphasis that the insights developed are biologically reasonable. We wish to obtain order of magnitude estimates for the set of parameters, $b$, $p_{x}$, $p_{u}$, $r$, $\delta$, $\kappa$, $R$, $m$ and $D$. We choose the yeast \textit{Saccharomyces cerevisiae} as our model organism. While our model is more physically realistic than many mathematical public good models, we note that there are still choices that must be made in relating this physical system to our general framework.

Our model is constructed such that the uptake of one constituent of the public good, $Q$, by a phenotype, results in a reproduction event. In the context of \textit{S. cerevisiae}, the type $Q$ is thus shorthand for the amount of invertase that must be present in the system to break down sucrose into sufficient glucose for a reproduction event of the yeast. Let us define $\sigma$ to be the scaling between $n_{q}$ and the total number of invertase molecules, such that the number of invertase molecules is $\sigma n_{q}$. In order to understand the relationship between our model parameters and physically measurable parameters, we begin by considering a simplified ODE system of our model.
\begin{eqnarray}
 \frac{ d x }{ d t } &=& x( b + r q - \kappa x ) \,,\\
 \frac{ d q }{ d t } &=& p x - \delta q \,.  
\end{eqnarray}
While the total number of discrete invertase constituents is $n_{q}\approx R^{2} q$, the total number of invertase molecules is $R^{2}\sigma q$. Let $\theta$ be a measure of the number of invertase molecules, such that $\theta=\sigma q$. The ODEs in this more natural variable read
\begin{eqnarray}
 \frac{ d x }{ d t } &=& x( b + \frac{r}{\sigma} \theta - \kappa x ) \,,\\
 \frac{ d \theta }{ d t } &=& \sigma p x - \delta \theta \,.  
\end{eqnarray}
The decay rate $\delta$ is independent of the number of molecules which make up an invertase constituent $Q$, so we can take experimental measurements of the invertase molecular decay rate as values for  $\delta$. Meanwhile the molecular invertase production rate and reproduction rate due to invertase take on scaled forms of the parameters in our original ODEs; 
\begin{eqnarray}
 r_{mol} &=& \frac{r}{\sigma} \,, \\
 p_{mol} &=& \sigma p \,.\label{eq:eq_pmol}
\end{eqnarray}

While measurements of $p_{mol}$ are obtainable in the literature (see Table \ref{tab:supmat_fig_parameter_exp}), our estimation of $r_{mol}$ is complicated by the fact that it is an effective parameter. It must capture the increase in the reproductive rate due to invertase, which in reality is coupled to both the reaction rate of invertase and sucrose into glucose, as well as the uptake rate of glucose by yeast and the energy conversion to reproduction. We do however know the typical range of yeast reproduction rates. Let us define $\lambda_{exp}$ as the yeast reproduction rate as measured experimentally. In turn, let $\lambda_{\text{eff}}$ be the effective per capita reproduction rate of yeast in the model;
\begin{eqnarray}
 \lambda_{\text{eff}} = b + r q \,.
\end{eqnarray}
The yeast reproduction rate clearly depends on the amount of public good in the system, typically varying from 
\begin{eqnarray}
 \lambda_{\text{eff}} &=& b \qquad\mathrm{(all\,non-producers)} \nonumber \\
 \mathrm{to} \qquad \lambda_{\text{eff}} &=& b \frac{  \delta \kappa }{ \delta \kappa - p_{x} r } \qquad \mathrm{(all\,producers)} \,.
\end{eqnarray}
In reality, the reproduction rate of yeast in a system without any invertase is effectively zero; we have assumed some baseline birth rate for convenience in the model, which could be physically interpreted as being associated with an exogenous glucose concentration in the system. We assume that this is typically low, such that $b$ is small, while the yeast approaches its maximum reproductive rate when it consists entirely of producers.

The parameter $\kappa$ controls death due to crowding. For simplicity this is the only form of death in the model. This choice leads, perhaps unnaturally, to the non-producers (who exist at typically lower densities) having a much smaller death rate than producers. For the parameters chosen however, we obtain per- capita death rates on the order of an hour for producers, and ten hours for non-producers. The parameter $R$ meanwhile measures the assumed spatial interaction scale. It determines the typical number of individuals on each patch. We can use this value to infer the size of a patch. Denoting the diameter of a yeast cell as $L_{c}$, and assuming that the hyper-producing cells in the stationary state can be packed on a grid, the size of each patch, $L_{p}$, can be approximated by
\begin{eqnarray}
 L_{p} 	&=& L_{c} \sqrt{ N_{u} } \,,\\
	&=& L_{c} R \sqrt{K_{u}} \,. \label{eq:eq_Lp}
\end{eqnarray}
The parameters $m$ and $D$ are effective migration and diffusion rates in our model. To map these physical parameters these must be in turn scaled by the patch length. The public good diffusion rate must also be scaled by $\sigma$, which maps the discrete amount of invertase constituents $Q$ to the number of invertase molecules. Denoting $m_{exp}$ and $D_{exp}$ the physical migration and public good diffusion rate rates, it can be shown that~\cite{gardiner_2009}
\begin{eqnarray}
 D_{exp} =  \sigma  L_{p}^{2}  D \,,\qquad m_{exp} =  L_{p}^{2}  m \,. \label{eq:eq_D_m_exp}
\end{eqnarray}
The parameter choices which follow from these calculations are summarized in \tref{tab:supmat_fig_parameter_list_1}.

\begin{figure}[t]
\setlength{\abovecaptionskip}{-2pt plus 3pt minus 2pt}
\begin{center}
\includegraphics[width=0.56\textwidth]{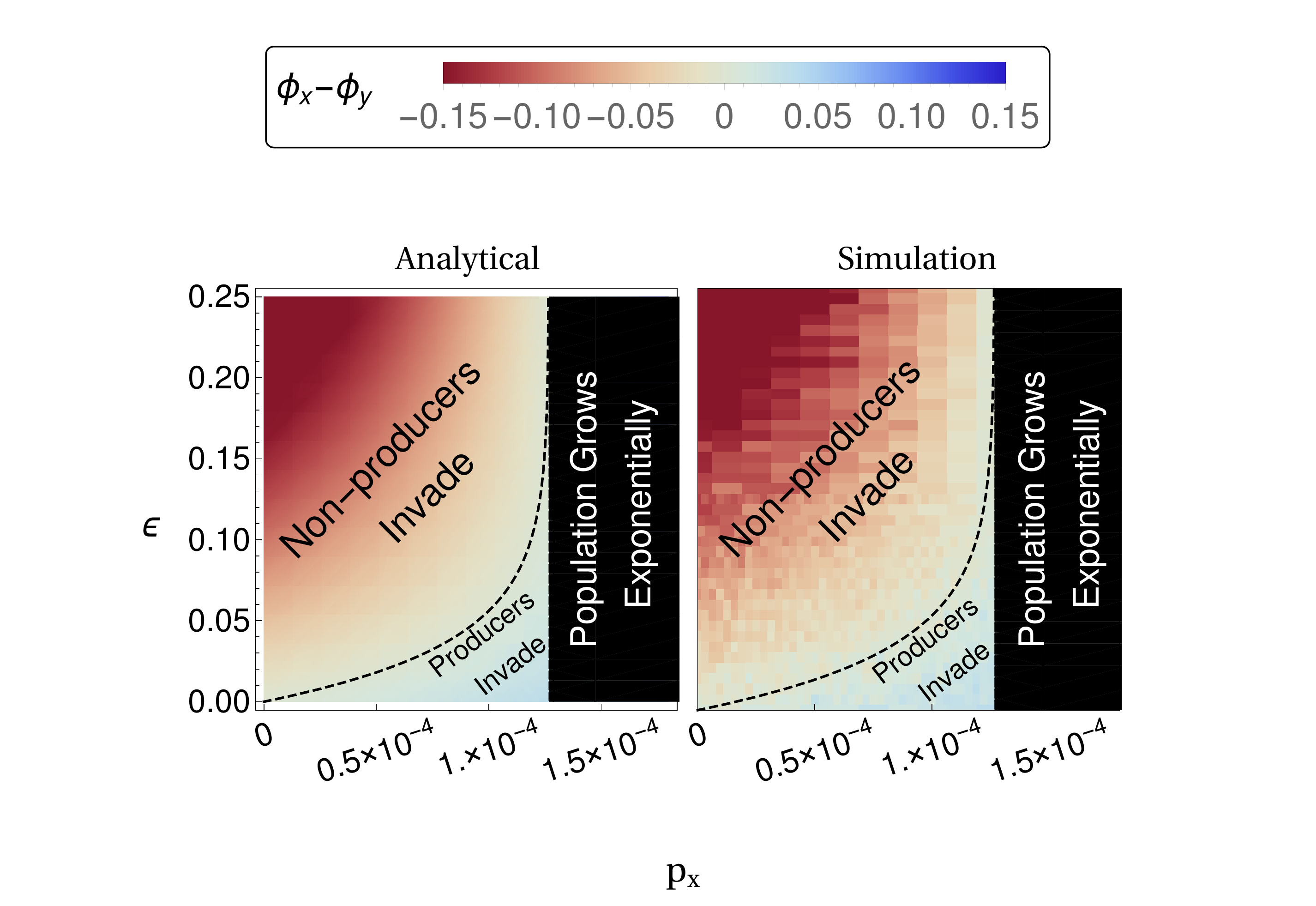}
\end{center}
\caption{Figure illustrating the larger range of values for the parameter $\varepsilon$ over which the approximation \eref{supmat_eq_PG_fixprob_f} is accurate. Parameters are given in  \tref{tab:supmat_fig_parameter_list_1}, with the exception of $p_{x}$ and $\varepsilon$ which are varied. Note that this figure is similar to Fig. 2, plotted in the main text, but plotted over a grater range of $\varepsilon$ and $\phi_{x}-\phi_{y}$. The parameter region plotted in black is that for which $ \kappa \delta > r p_{x} $.}\label{supmat_fig_trade_off_larger_eps}
\end{figure}

When considering the parameter choices summarized in \tref{tab:supmat_fig_parameter_list_1}, it is important to make a final point. While both the approximations we have employed, the system size expansion in \sref{sec:supmat_sec_obtain_SDE_system} and the fast-variable elimination in \sref{supmat_sec_public_good_reduction}, rely formally on $R^{2}$ being large and $\varepsilon$ being small ($\mathcal{O}(R^{-2}) > \varepsilon$), in practical terms the procedures are relatively robust to this restriction. Indeed, throughout the body of the main text, $R=4$, while $\varepsilon$ is varied on the interval $[0,0.1]$. In fact we find that the approximate analytic expression we obtain for the invasion probabilities of the phenotypes, \eref{supmat_eq_PG_fixprob_f}, describes the results obtained from simulation well, up to $\varepsilon=R^{-2}$, as illustrated in \fref{supmat_fig_trade_off_larger_eps}. 

In terms of the system size expansion, this robustness can in part be explained by the fact that the typical population sizes ($N_{x}$, $N_{y}$ and $N_{q}$) are proportional to $R$. For populations of fixed size $N$, SDEs for the system can be obtained by means of a Taylor expansion of the master equation (for example, \eref{eq:supmat_meqn_general}) as a series in $1/N$. A crucial feature of the system we are concerned with here however, is that population sizes may vary, and so this technique is unavailable. Instead we conduct an expansion in the interaction scale $R$, which is proportional to the mean population size. Though $R$ may not be a large number itself, increasing $R$ leads to an associated increase in population size (see \tref{tab:supmat_fig_parameter_list_1}). In turn, this leads to terms of higher order in the Taylor expansion of the master equation becoming subdominant~\cite{van_kampen_2007}, justifying the truncation which leads to \eref{eq:supmat_eq_FPE_general}. In contrast, the resilience of the fast-variable elimination approximation to such large values of $\varepsilon$ is surprising.

\onecolumngrid

\begin{table}[h]
\setlength{\abovecaptionskip}{-2pt plus 3pt minus 2pt}
\begin{center}
\begin{tabular}{ |l|l|l| }
\hline
\multicolumn{3}{ |c| }{ Experimental parameters }  \\
\hline
Experimental Parameter 	& Value 						& Description \\ \hline
 $p_{mol}$		& $0.46 \, \mathrm{mol}\,s^{-1}$			& Production rate of a molecule of invertase per \\
			&							& producing yeast cell~\cite{koschwanez_2011}.  	\\ \hline
 $\delta$		& $2 \times 10^{-3} \, \mathrm{mol}\,s^{-1}$		& Estimated efficacy decay rate of invertase (see \cite{gomez_2008}, Fig 5.)	\\ \hline
 $\lambda_{exp}$	& $0.31-0.5 \, hr^{-1}$ 				& Yeast reproduction rate in producing population~\cite{sanchez_2013,snoep_2009}  	\\ \hline
 $\varepsilon_{exp}$	& $0.06$						& Cost of public good production to yeast reproduction rate~\cite{sanchez_2013}		\\ \hline
 $D_{exp}$		& $100\, \mu m^{2} s^{-1}$ 				& Diffusion rate of invertase molecules estimated in \cite{allen_2013}  	\\ \hline
 $L_{c}$		& $3\,\mu m$						& Cell length physical approximation \cite{allen_2013}.		\\ \hline
\end{tabular}
\end{center}
\caption{List of experimental parameters obtained from literature.}\label{tab:supmat_fig_parameter_exp}
\end{table} 

\settablecounter{8}{2}
\begin{table}[h]
\setlength{\abovecaptionskip}{-2pt plus 3pt minus 2pt}
\begin{center}
\begin{tabular}{ |l|l|l| }
\hline
\multicolumn{3}{ |c| }{ Illustrative parameter choices with justifications }  \\
\hline
      Parameter & Value 							& Justification \\ \hline
 $\sigma$	&	$4000$							& Assumed parameter. Presence of $4000$ invertase  \\
		&								& molecules required for yeast reproduction.\\ \hline
 $p_{y}$	& 	0	 						& True non-producer does not produce invertase.			\\ \hline
 $p_{x}$	&    	$1.14\times10^{-4}\,s^{-1} $, 				& Experimental value of molecular invertase production 		\\ 
		&	($0.41\,hr^{-1}$)					& rate (see Table \ref{tab:supmat_fig_parameter_exp}) scaled by $\sigma$ (see \eref{eq:eq_pmol}).	\\ \hline
 $p_{u}$	&    	$ 1.2 \times 10^{-4} \,s^{-1} $			 	& Leads to factor $1.7$ increase in the steady state invertase 	\\ 
		&	($0.43\,hr^{-1}$)					& from producing to hyper-producing population, consistent with \cite{maclean_2008}.	\\ \hline
 $b$		&	$6.94\times10^{-6}\,s^{-1}$				& Small baseline yeast birthrate assumed.  	\\ 
		&	($0.025\,hr^{-1}$)					& 	\\ \hline
 $r$		&	$ 1.58 \times 10^{-5}\, s^{-1} $			& Chosen so as to give per-capita yeast reproduction rate \\ 
		&       ($0.057\,hr^{-1}$)     					& $( b + r q) \approx \lambda_{exp}$ when system entirely producers (see Table \ref{tab:supmat_fig_parameter_exp}).\\ \hline
 $\delta$	&	$0.002\,s^{-1}$						& Taken from experimentally measured values. (see Table \ref{tab:supmat_fig_parameter_exp}) 		\\ 	 \hline
 $\kappa$	&    	$1 \times 10^{-6}\,s^{-1}$				& Suggested parameter for illustrating effects in paper; \\ 
		&								& restricted by $\delta \kappa > p_{i} r $, $i=x,y,u$.	\\ \hline
 $R$		&	$ 2 $							& Suggested parameter for illustrating effects in paper.	\\ \hline
 $\varepsilon$	&	$0.06$							& Taken from experiments (see Table \ref{tab:supmat_fig_parameter_exp}) \\ \hline
 $N_{y}$	& 	$ 28 $							&  See \eref{eq:supmat_eq_NX_NY}.	 \\ \hline
 $N_{x}$	& 	$ 302 $							&  See \eref{eq:supmat_eq_NX_NY}.	 \\ \hline
 $N_{u}$	&	$ 499 $							&  See \eref{eq:supmat_eq_NX_NY} for $N_{x}$ and substitute $p_{u}$ for $p_{x}$.	\\ \hline
 $L_{p}$	&	$67 \,\mu m$						&  See \eref{eq:eq_Lp}.							\\ \hline 
 $m$		&	$3.4\times10^{-7} \, s^{-1}$				&  Yields a migration to birth-rate ratio between $m/b=4.9\times10^{-2}$ \\
		&								&  (all non-producers) and  $m/(b+r q)=4.5\times10^{-3}$ (all producers).   \\ \hline
 $D$		&	$2.22 \times 10^{-5} \, s^{-1}$				&  Obtained using experimental value $D_{exp}$ from Table \ref{tab:supmat_fig_parameter_exp} and \\
		&								& \eref{eq:eq_D_m_exp}. \\ \hline 
\end{tabular}
\end{center}
\caption{List of parameters used in the simulation, with the exception of $p_{x}$, $p_{u}$, $\varepsilon$, $m$ and $D$ which are varied.}\label{tab:supmat_fig_parameter_list_1}
\end{table}

\twocolumngrid

\clearpage

\section{Movies}\label{sec:supmat_sec_movies}

\setcounter{figure}{0}
\begin{figure}[h]
\renewcommand{\figurename}{Movie}
\setlength{\abovecaptionskip}{-2pt plus 3pt minus 2pt}
\begin{center}
\includegraphics[width=0.45\textwidth]{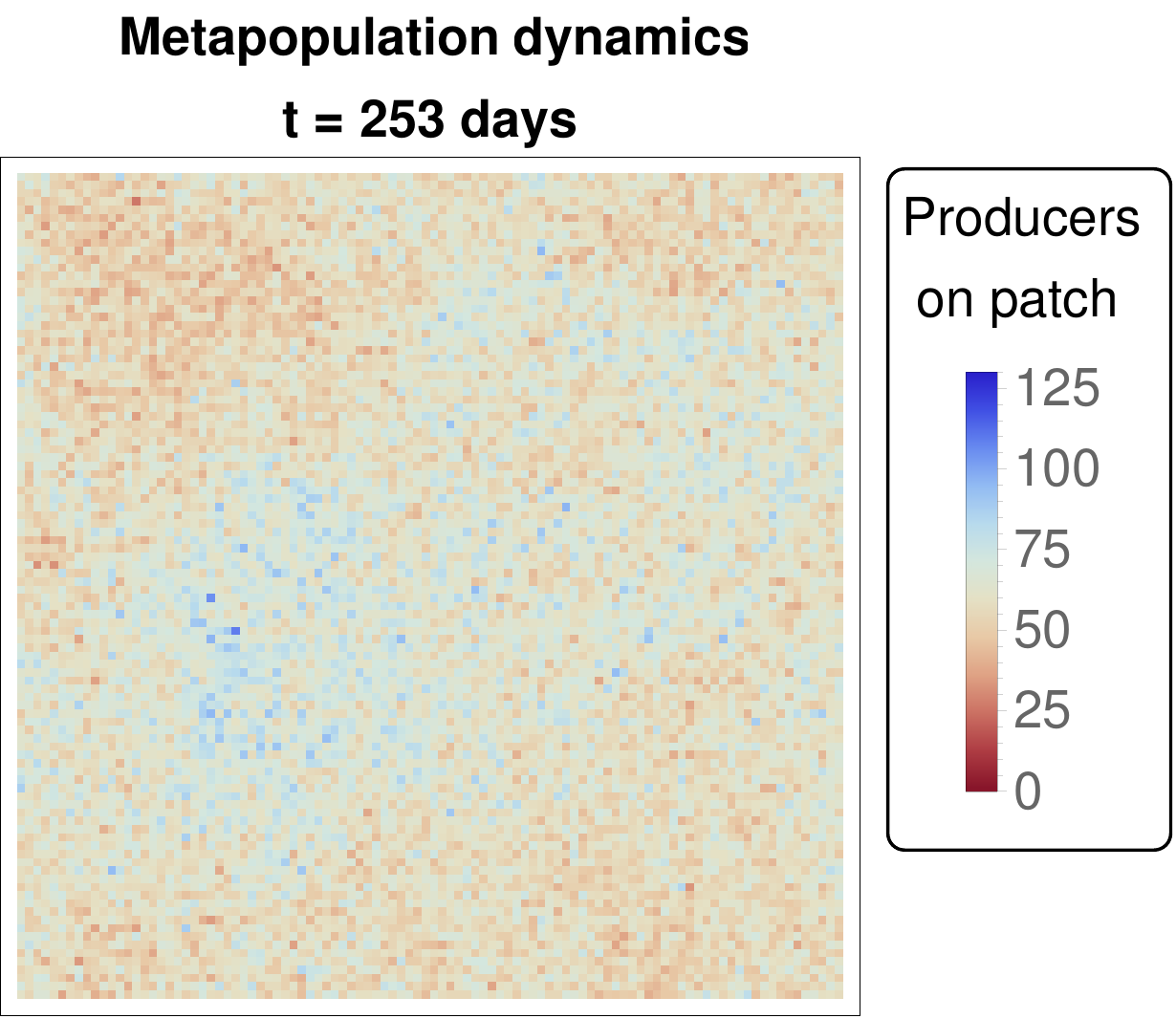}
\end{center}
\caption{Still from Movie S1: Movie of a simulation of the metapopulation public good model on a $100\times100$ grid ($C=10^{4}$). Colors indicate the number of producers on each patch; patches with a small number of producers are colored red while patches with a large number of producers are colored blue. Parameters used are $p_{x}=1\times10^{-4}$, $\varepsilon=0.02$, $m=3.7\times10^{-5}$ and the remaining parameters taken from \tref{tab:supmat_fig_parameter_list_1}. With these parameters, $N_{y}\approx28$ and $N_{x}\approx129$. Initial conditions are a single producer and non-producer on each patch. Large numbers of producers on a patch are correlated with low numbers of non-producers on the same patch. The space-averaged dynamics of this simulation are given in the main text, Fig.\,4. Counter to the deterministic prediction, the number of producers increases with time, while the number of non-producers decreases.}\label{supmat_fig_video_S1}
\end{figure}

\begin{figure}[h]
\renewcommand{\figurename}{Movie}
\setlength{\abovecaptionskip}{-2pt plus 3pt minus 2pt}
\begin{center}
\includegraphics[width=0.45\textwidth]{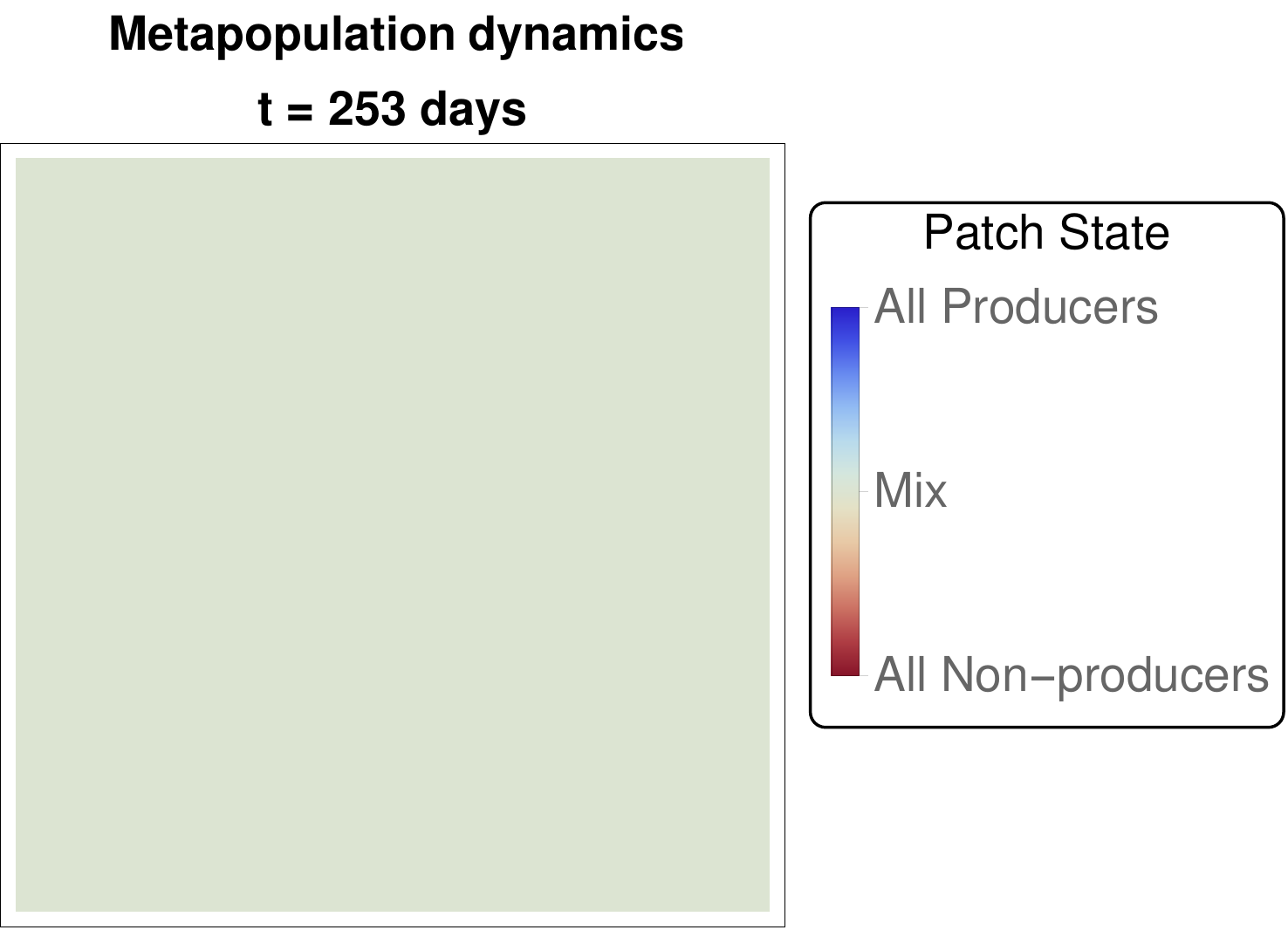}
\end{center}
\caption{Still from Movie S2: Movie showing the distribution of homogeneous non-producing patches (red), homogeneous producing patches (blue) and heterogeneous mixed patches (gray-green) in the simulation of the metapopulation public good model given in Video S1. For the majority of the observation time, every patch contains a heterogeneous mix of producers and non-producers. Homogeneous producer patches only begin to emerge as producers approach fixation in the system.}\label{supmat_fig_video_S2}
\end{figure}

\clearpage


\end{document}